\documentclass[iop, apj]{emulateapj}     

\begin{document}
\topmargin=0.1in


\title{``Observing and Analyzing'' Images From a Simulated High Redshift Universe  }

\author{Robert J. Morgan, Rogier A. Windhorst, Evan Scannapieco}
\affil{School of Earth and Space Exploration, Arizona State University, Tempe, AZ 85281-1406}
\author{and Robert J. Thacker }
\affil{Dept. of Physics and Astronomy, St. Mary's University, Halifax, NS B3H 3C3, Canada}

\journalinfo{{ \rm Accepted for publication in} the Publications of the Astronomical Society of the Pacific} 

\begin{abstract}

We investigate the high-redshift evolution of the restframe UV--luminosity 
function (LF) of galaxies via hydrodynamical cosmological simulations, 
coupled with an emulated observational astronomy pipeline 
that provides a direct comparison with observations.
We do this by creating mock images and synthetic galaxy catalogs 
of $\mathrm{ \approx 100 \ arcmin^2 }$ fields 
from the numerical model at redshifts $\approx 4.5 $ to 10.4.
We include the effects of dust extinction and the point spread function 
(PSF) for the Hubble WFC3 camera for comparison with space observations. 
We also include the expected zodiacal background to predict its effect on 
space observations, including future missions such as the James Webb Space 
Telescope (JWST). When our model catalogs are fitted to Schechter function 
parameters, we predict that the faint-end slope ($\alpha$) of the LF evolves 
as ${\alpha = -1.16 - 0.12 \ z}$ over the redshift range z $\approx 4.5 $ 
to 7.7, in excellent agreement with observations from e.g., Hathi et al. 
(2010). However, for redshifts $ z \approx 6$ to 10.4, $\alpha (z)$ appears 
to display a shallower evolution, ${\alpha = -1.79 - 0.03 \ z}$.
Augmenting the simulations with more detailed physics --- specifically stellar 
winds and supernovae (SN) --- produces similar results.
The model shows an overproduction of galaxies, especially at faint magnitudes, 
compared with the observations, although the discrepancy is reduced when 
dust extinction is taken into account.

\end{abstract}


\clearpage

\section{Introduction}
The luminosity function (LF) of galaxies is an important indicator of galaxy
assembly and evolution (Schechter, 1976). It traces the star-formation rate 
(SFR) and contains clues to physical processes such as galaxy merger rates,
and also feedback from supernovae (SN) and active galactic nuclei (AGN). 
Numerical simulations are essential to understanding and predicting
the highly non-linear processes in large scale structure formation, but
are often difficult to interpret with respect to the actual astronomical 
observations. This is due to many factors, such as the presence of detector 
noise and the sky-background. It is thus non-trivial to draw precise 
inferences from a comparison between numerical predictions 
and the images actually observed in astronomy.  Overzier et al. (2013) 
characterized the current state of the relation between theory 
and observation as being mostly one-directional, in that ``physical quantities 
estimated from observations were compared with theoretical predictions,''
and that a greater understanding could be gained by going in the other 
direction.

To help bridge this gap between theory and observations, we extend the 
numerical models to include the simulation of an observational pipeline. 
That is, we form ``artificial'' images from the simulations of galaxy 
formation, using the stellar population synthesis models of Bruzual \& Charlot 
(2003, hereafter BC03) to predict the observed stellar light distribution, 
and process these to create mock galaxy catalogs using Source Extractor 
(``SExtractor''; Bertin \& Arnouts 1996) after imposing the appropriate 
amount of image noise due to the zodiacal background. 

Overzier et al. (2013) summarized previous work done in this area, 
such as Bouwens et al. (1999), who coupled galaxy evolution 
models to spectral energy distribution (SED) models to predict luminosities at 
high redshifts, and Bouwens, Illingworth \& Magee (2006), who predicted the
evolution of galaxy properties by artificially redshifting selected galaxy
images, producing new synthetic images. They also discuss producing 
``observed'' image sets from semi-analytic models (SAMs).

Realistic synthetic telescope images produced from the output of
hierarchical simulations to emulate sky surveys were generated by e.g.,
Blaizot et al. (2005). Synthetic observations are also produced using the 
output of SAMs by processing the results of DM halo simulations. They used 
methods similar to those used here, in terms of the use of translation and 
rotation of multiple slices of data at different redshifts to produce 
artificial images and catalogs. 

There have also been many numerical simulations of galaxies without 
producing mock observations, mostly with SAMs (e.g., Bolton \& Haehnelt 2007; 
Mao et al. 2007; Samui et al. 2009; Trenti et al. 2010; Dayal et al. 2014), 
as well as Overzier et al (2013). Dayal et al. (2014) used models of 
star-formation and merger trees of DM haloes to simulate properties of high 
redshift galaxies. Hydrodynamical simulations with dark matter (DM), gas and 
stars have been used to study galaxies (e.g., Jonsson et al. 2006; Lotz et al. 
2008, 2010; Robertson \& Bullock 2008; Wuyts et al. 2009; and Chiliangarian et 
al. 2010). However, these were mainly used to study the effects of dust on the 
observations, and were not on cosmological scales.  Relatively recently, 
there have been hydrodynamical simulations, investigating the galactic LF 
at high redshift (e.g., Nagamine et al. 2006; Finlator, Dav\'{e}, \& 
Oppenheimer 2007; Salvaterra, Ferrara \& Dayal 2011; Gabor \& Dav\'{e} 2012, 
Jaacks et al.  2012; Stinson et al. 2013; Shimizu et al. 2014).

Jaacks et al. (2012) included star--formation, dust extinction plus IGM 
transmission effects. They calculated the SED of each selected galaxy using 
BC03 to make spectrophotometric comparisons with the observations.  They also 
also ran different simulations at different resolution scales to cover the 
bright, medium and faint-end of the LF scale, and then combined the results. 
We discuss this later in $\S{5.4}$.

Salvaterra et al. (2011) have also studied the high redshift galactic LF.
Shimizu et al. (2014) performed large hydrodynamical simulations with both
DM and baryonic matter including star--formation, SN feedback, and feedback 
winds. They calculated the SEDs of each star particle and included dust 
attenuation, and created mock observations in order to make comparisons with 
the observed UDF12 field. We discuss the results of these groups later in 
$\S{6}$, when discussing our own results.

In this work, we use gas and hydrodynamic models based on the Gadget2 code 
(Springel (2000); Springel et al. 2001; Springel \& Hernquist  2003), 
including star-formation, to avoid many of the assumptions necessarily
made in SAM approaches. The use of hydrodynamic numerical simulations 
limits our ability to reproduce survey-size scales of many square degrees 
over cosmological time scales. However, we are able to produce images on 
the scales of the Great Observatories Origins Deep Surveys (GOODS) mosaics 
fields of 8 -- 10 arcminutes with HST, and in the future also with JWST, at 
redshifts of $z \simeq 4.5$ to 11. It is time-consuming to run hydrodynamical 
simulations with different sets of parameter values. Hence, it is crucial 
to perform these simulations with minimal assumptions, to see how well we 
can reproduce the actual observations with physics-based model inputs. This 
also aids in deducing the possible effects of different physics on the outcome, 
as these can be added to the model.  This was also done by Haas (2010), who 
used hydrodynamic simulations from the OverWhelmingly Large Simulations
(Schaye et al. 2010) to study the LFs of mock galaxies, though this work
was in the $K$-- and $B$--  band filters and confined to lower redshifts, 
$z  <  4$.

Overzier et al. (2013) pursued a similar approach in the ``Millennium Run
Observatory'' (MRObs), using semi-analytical models. They implemented a rich 
user-interface for perusing the simulation data-base at different 
wavelengths and filters, and included a number of observational artifacts.  
Our work is more focussed on predicting the faint-end of the luminosity 
function (LF) of galaxies, something that is of considerable interest to 
observers and relevant to the direct planning of missions such as JWST. The 
results presented here can thus potentially be used to guide the planning of 
future JWST observations.

\section{Methodology}

\subsection{Summary of Method}


Numerical simulations are used to predict synthetic images
similar to those that may be observed at high redshifts ($\S{2.2}$).
Our simulation code is a variant of Gadget2 (Springel 2005)
that includes additional physics modules, such as star-formation and
feedback from supernovae as well as radiative cooling processes.

Since our intent was not to create an exact simulation
of any particular instrument --- but rather to include the most important
and general observational artifacts, including adding the effects from dust --- 
some simplifications were made. The photon energy is considered to be solely 
from the simulated stellar populations, without significant reprocessing.
Hence, in our first analysis we do not consider extinction by dust, although
we discuss its effects later (\S{3.6}). There is also no attempt to include 
artifacts such as confusion from foreground galaxies, whether from natural 
overlap (Windhorst et al. 2008) or gravitational lensing (Wyithe et al. 2011). 
Note that confusion from foreground stars can be safely ignored, as the faint 
star-counts have a much flatter slope than the faint galaxy counts in the 
red-near-IR (e.g., Windhorst et al. 2011).

Simulation outputs at various redshifts include stellar particles that model 
the star-formation and evolution in the simulation, as explained in $\S{3.1}$. 
We treat these particles as simple stellar populations (SSPs) with a given age 
and metallicity, as determined in the model. For these, spectral energy 
distributions (SEDs) in the emitted rest-frame are derived using 
the BC03 models ($\S{3.2}$). Fluxes are calculated in the simulated observer's 
frame, integrated with filter response functions ($\S{3.3}$), and converted to 
images ($\S{3.1}$) in the FITS (Greisen et al. 1980) format ($\S{3.4}$). 
As explained in $\S{3.1}$, the size of each frame in pixels is determined
from the relative comoving distance in our adopted $\Lambda$CDM cosmology.
This permits ``stacking'' of frames at different redshifts. 

We assume WMAP values available at the time of the simulation, 
$\mathrm{H_{0} = 71.9}$, $\Omega_{\Lambda}$ = 0.742, and $\Omega_{b} =0.0441$,
(Komatsu et al. 2009). A slightly different value (0.73) for $\Omega_{\Lambda}$
was used for the calculation of distances, due to more recent WMAP7 values 
(Komatsu et al.  2011). However, this does not affect the restframe absolute 
magnitude calculation, since the same luminosity distance was used in 
calculating the flux in the observer's rest-frame.
Since the simulations were run, more recent values have become available
from Planck (2015), but as our values are within $\sim 5\%$ of those, this 
shouldn't make much of a difference, since the $\emph{simulated physical}$ 
parameters are likely uncertain by at least this much.

The method of projecting the 3-D data onto the 2-D sky-plane is described
in $\S{3.1}$.  These images are analyzed by the SExtractor package 
(Bertin \& Arnouts 1996) to find ``SF-particle'' close groupings (see Fig. 1), 
that are treated as `galaxies' or `galaxy building blocks', and binned into 
observed magnitude ranges. This process is then repeated, adding effects from 
dust extinction and PSF-convolution ($\S{3.6}$).

The luminosity functions (LFs) are fitted to a Schechter (1976) function,
and best-fit estimates of the faint-end LF slope parameters
$\alpha$, characteristic magnitude $M^*$, and the density normalization 
$\phi$* are found using chi-square minimization ($\S{4}$). Contours of 
chi-square in the $\alpha - M^*$ plane for best-fit normalizations, 
as well as the LF slope $\alpha$, are presented in $\S{5}$ as a function of 
redshift and compared with the observations from Hathi et al. (2010).

We repeat the analysis for different model parameters intended to include
effects from stellar and galactic winds. Also, the appropriate amount of
simulated `sky-noise' (foreground zodiacal light; for a summary, see Table 2
of Windhorst, et al. 2011) is added to the images to 
simulate deep-space environments, and the image extraction and data reduction 
process is repeated to evaluate the effect of these noise sources on the 
``observation'' of the underlying SPH model. This process is also repeated 
for adding effects from dust extinction and PSF convolution ($\S{3.6}$).

We note that this use of SExtractor is a departure from the usual method of 
treating simulated galaxy data, but not actual observed galaxy images. The
intent of this work is to create ``mock observations'' using ``mock galaxy
images''.  To check this use of SExtractor, a comparison was made with a
publicly available ``friend-of-friends'' (``FOF'') tool from the U. of 
Washington Dept. of Astronomy 
\footnote{http://www-hpcc.astro.washington.edu/tipsy/top.html}, 
which is a more usual method of extracting groupings from this type of 
simulated data. We found a 96.9 percent correlation between the number of 
groups found using the two methods on the same simulated frame. 
Specifically, using a 20 Kpc ``linking length'' parameter on a snapshot at 
redshift 7.3, ``FOF'' found 8987 groups, while SExtractor found 8711 objects. 
We attribute the discrepancy to SExtractor's requirement that pixels with 
flux above the threshhold be contiguous in order to be considered an object. 
This agreement is sufficiently good to proceed with the method.

\subsection{Details of the Numerical Simulation}

The numerical simulations include both dark matter and gas particles. 
Besides simulating the effects of gravity, it uses smoothed 
particle hydrodynamics (SPH), (Gingold \& Monoghan 1977; Lucy 1977), 
heating and cooling (Katz et al. 1996), and star-formation (Springel 
\& Hernquist 2003, hereafter SH03). A hybrid gas/star-particle model is 
employed, which has both cool and hot gas, and stellar components for sub-grid 
modeling. Star `particles' are formed from the hybrid gas-particles, 
representing a stellar population or star-cluster, when 50\% of the hybrid 
particle has been converted to stars.  

While the exact details of the star-formation procedure are complex 
(see SH03 for a detailed description), the rate of production of stars
is essentially determined by the density of cold gas clouds divided by
the characteristic timescale of star-formation, $\tau$. An external
UV-background, based upon a modified Haardt \& Madau (1996) spectrum is
also included (see Dav\'{e} et al. 1999 for details). This background turns on
at $z\approx6$ and is included in the calculation of ionization state 
abundances of H and He, which determines the net cooling rate and thus
impacts the formation of cold clouds.

The simulation volume was a cube 18 comoving Mpc/$h$ on a side,  
($h\mathrm{=H_{0}/100 \ km \ sec^{-1} \ Mpc^{-1}}$), with periodic boundary 
conditions, using $2 \times 512^{3}$ particles, with equal numbers of dark 
matter and gas particles, 
yielding a baryon-mass resolution of $\mathrm{5.4 \times 10^{5}}$ ${h^{-1}\ }$ 
M $_{\odot}$ per particle. 

Initial conditions were generated using 2nd order Lagrangian perturbation theory
(e.g., Thacker \& Couchman 2006) at an initial redshift of z = 199.
Model outputs of particle data were recorded at time intervals
corresponding to two light crossing times of the simulation cube, where
a light crossing time is the time it takes light to travel from one side to
the opposite side, ignoring the small expansion of space in this time. Thus,
the light crossing time is given by 
$\mathrm{18\ Mpc \ }$  $h^{-1} (1+ $  z$ )^{-1} $c$^{-1}$.
The star-formation code creates stars in the multi-phase gas particles
when a sufficient fraction of the gas reaches appropriate temperature and 
density conditions (SH03). When 50\% of the hybrid particle mass is processed 
into stars, that portion is split off as a ``star-particle.'' These 
star-particles have masses $\mathrm{ 2.7 \times 10^{5} }$ $h^{-1}\ $M$_{\odot}$,
or roughly a globular cluster or Giant Molecular Cloud (GMC) mass.
The particle files were processed to extract data on the
star-particles, including mass, metallicity and formation time, 
which were used to obtain rest-frame SEDs from the BC03 models.
In later runs, parameters were set to permit the simulation of feedback via
galactic winds, largely from simulated supernovae, and shocks and heating of 
the gas from the SNe and star-formation. These runs are described as including 
`winds' in the text and figures.  See Table 1 for parameter values, which
were taken from SH03.

\begin{table}[ht]
\caption[]{Cosmological and Physical Simulation Parameters Used.} 
\begin{tabular}{l l}\hline
\multicolumn{1}{c}{ Parameter} &
\multicolumn{1}{c}{Value} \\ \hline \hline
OmegaLambda ($\Omega_{\Lambda}$) & 0.742  \\\hline
OmegaBaryon ($\Omega_{B}$) & 0.0441  \\\hline
HubbleParam (h) & 0.719  \\\hline
Softening Length & 1.8 h$^{-1}$ kpc   \\ \hline \hline 
\multicolumn{2}{c}{`Winds' Parameters} \\ \hline \hline
WindEfficiency &  0.5  \\\hline
WindEnergyFraction &  0.25  \\\hline
WindFreeTravelLength &  0.5  \\\hline
\end{tabular}
\newline
\newline

Parameters used in simulations 
\emph{without} feedback ``winds'' and \emph{with}. Names used are the 
parameter names in the model, except for `Softening Length', which is the 
gravitational softening length and the SPH kernel size. Feedback `Wind' 
parameter values were taken from SH03.  Lengths are in comoving units.

\end{table}

\begin{table}[ht]
\caption{Simulated SExtractor Object Counts.}
\begin{tabular}{rrrrr}\hline
\multicolumn{4}{c} {} \\ %
Redshift & No BG & Sky BG & Recovered & Number of \\ 
 &  &  & Fraction & Snapshots  \\\hline
4.52 & 15334 &  2407 & 0.157 & 1\\\hline
5.32 & 13809 & 2140 & 0.155 & 1\\\hline

6.01 & 13392 & 1986 & 0.148 & 1\\\hline
6.24 & 49818 & 7235 & 0.145 & 5\\\hline
7.16 & 39766 & 3223 & 0.081  & 5\\\hline
7.68 & 36213 & 3248 &  0.090 & 5\\\hline
10.38 & 12702 & 967 & 0.076 & 5\\\hline
\end{tabular}
\newline
\newline

Recovered Object fraction ``with'' compared to ``without'' sky 
background (BG).
\newline
SExtractor counts of the same synthetic galaxy fields at different
redshifts without added sky BG, and with added sky BG
$\mathrm{\approx 22.6 \ AB}$-$\mathrm{mag \ arcsec^{-2}}$. 
At redshifts  $z \geq 6.24$, five (5) snapshots are combined, while at 
$ z < 6.24$ only a single snapshot is used, hence the increase in counts. 
For details, see text.
We note a sharp decrease in completeness (ratio of counts with sky BG 
to counts without sky BG) at redshift $z$ $\geq 7.16$.
We also note a sharp drop in the total counts for redshift $z >$ 7.68. 
\end{table}

\section{Image Synthesis}
\subsection{Simulated Image and Pixel Scales}
%
The simulated image pixel-scale was chosen to be comparable to the
resolution of space-based instruments such as HST and JWST.
This scale is also consistent with the size of astrophysical objects 
most closely represented by the model particles, such as GMCs  and massive  
star-forming regions with masses  $\mathrm{M \leq 10^6 \ {M}_{\odot}}$.
That is, the hybrid gas-star particles represent star-forming regions.
The pixel-scale of the WFC3 camera on the HST at the near-IR wavelengths 
simulated here is 0.13"/pixel (Windhorst et al. 2011.) JWST will have a 
near-IR pixel scale of 0.034 -- 0.068", with a near-IR image resolution of 
$\approx 0.06"$ FWHM (Gardner et al. 2006).

A reference scale was set at redshift $z = 3.0$ for the FITS frame sizes. 
For the chosen $\Lambda$CDM cosmology, a redshift $z \simeq 3$ corresponds to 
a comoving distance of 6.4 Gpc.  A sky-field of 18 comoving Mpc/h on a side 
would then be 808 $\times$ 808 arcsecs, roughly the size of GOODS-sized 
mosaics with HST, and similar future projects that the community will likely
propose for JWST. Specifically, the JWST NIRCam --- which covers the range of 
0.6 to 5.1 microns --- has a field of 132 $\times$ 264 arcsecs 
\footnote{http://www.stsci.edu/jwst/instruments/nircam/operations\#imaging}.
A typical mosaic is expected to be $\approx 3 \times 4$ tiles,
or $\approx 400 \times 500$ arcsecs, which would be of the order of the
higher redshift simulated images, since the size scales inversely with the
comoving distance.

One of the reasons for doing our simulations is to get better guidance how 
to best plan searches for First Light objects at z$\gtrsim$10.
Currently, the JWST Guaranteed Time Observations (GTO) observations are 
planned to be a combination of 1--2 deep fields (1--2$\times$100 hrs) to
AB$\lesssim$31 mag, plus a larger number ($\sim$10$\times$10 hrs) of 
complementary medium-deep fields to AB$\lesssim$30 mag. In addition, we 
anticipate that the JWST GO community may propose for one JWST
UltraDeep Field (1$\times$800 hrs) to AB$\lesssim$32 mag, and possibly
also a set of CANDELS or COSMOS like ultra-wide JWST fields to
AB$\lesssim$29 mag. Combined, these wedding-cake layered JWST GO and GTO
surveys would in the end provide the best combination of depth, dynamic
range, area, and sampling of cosmic variance of the LF at
4$\lesssim z\lesssim$11 or higher. Our simulations are thus designed to
cover at least the observed flux range of 24$\lesssim$AB$\lesssim$31 mag
in such JWST surveys. 

We chose an image size with a power of 2, to conform to common detection 
formats.  A field of 8192 $\times$ 8192 pixels yields a scale of 
0.099"/pixel, which is in between that of Hubble WFC3/IR (0.13"/pix) and 
JWST NIRCam (0.034 to 0.068"/pix).  As the simulation has a fixed volume in 
comoving space, the simulated frame-sizes are inversely proportional to the 
comoving distance, which also maintains the pixel angular scale. 

While the simulation does not permit pixel-resolution down to actual 
astrophysical objects with masses less than $\mathrm{ 10^{5}\ M_{\odot}}$
--- although subgrid methods are used to simulate smaller sizes --- 
this still corresponds roughly to the upper end of GMC and massive 
star-forming region masses. These have typical sizes of order 
50 --- 100 pc. At $z=4.5$, 100 pc corresponds to 0.015", at $z=6$, to 0.017", 
and at $z=10$, to about 0.024".  Thus, a $\mathrm{\sim 10^5\ M_{\odot}}$ 
object at this pixel scale would be sub-pixel at these distances, even 
for JWST (note that these are proper sizes here). 
At redshifts higher than 6.3, the simulated frames were formed by
``stacking'' five (5) consecutive snapshots, in order to simulate a larger 
volume of the simulated ``observed'' space. Table 2 shows the  number of
snapshots used for each redshift image. To avoid aliasing effects --- since 
the same space was being captured at different lookback times --- advantage 
was taken of the simulation's periodic boundary conditions.  Each frame was 
rotated in a cyclic permutation about the edges of the data cube before 
being projected on the sky-plane corresponding to the new $x-y$ plane. Next, 
the images were shifted randomly in the sky-plane, applying periodic conditions.
These projections were then combined into a single frame. The effective 
redshift was taken as the median of the contributing frame-redshifts. 
The size of the merged frames was taken as the minimum (farthest) image 
frame in the contributing set of that redshift slice.

The total volume was computed by truncating the closer (or lower redshift) 
volumes, since this method included only the solid angle of the furthest frame.
Successive ``snapshot'' files were written at double the light-crossing time 
for the simulation volume. Hence, successive frames are quite close in their 
corresponding redshifts.  We typically have redshift slices spaced by 
$\sim$ 8--15 Myr in cosmic time.  Single time-slices were chosen for the 
redshift bins for $z \le 6.3$, and 5 time-slices were combined, or ``stacked'', 
for $z > 6.3$ to keep the resulting database manageable. 

\subsection{Interface to the BC03 models}

The simulation output binary-files were read, and new files written from 
these data for the star-particles, including model age and metallicity.
The age of the star-particle was determined by when it was formed relative to
the cosmic time of the model output. The cosmic age of the output data was 
computed from the redshift of the time-slice, according to the $\Lambda$CDM 
model assumed, although the value of $\Omega_{\Lambda}$ was updated to 0.73 
to reflect more recent data, as explained earlier. The metallicity and age 
were used to map to the appropriate stellar population model and SED in BC03, 
using the Chabrier (2003) IMF models. We chose the 150 nm restframe region 
for comparison with the observations in Hathi et al. (2010). These SEDs were 
redshifted to the (virtual) observer's rest-frame, and convolved with a filter 
response function, then multiplied by a simulated telescope aperture and 
exposure time. These were recorded in a file, along with particle comoving 
coordinates and model age and metallicity data.

Filters were chosen to conform to the criteria in Dahlen et al. (2007).
This required that the redshifted 150nm emitted-wavelength lie between the 
25th and 75th percentiles of the filter's response function. Filters $i$ and 
$z$ (Gunn), $J$ (Johnson), and $H$ (Bessel and Brett) were taken from the 
available filters in the BC03 filter file, but the F105W filter from HST 
WFC3 was added to the file (i.e., Windhorst et al. 2011, Koekemoer et al. 2011).

The integrated flux from the BC03 models was multiplied by the star-particle 
mass in $\mathrm{M_{\odot}}$, and converted to ergs/sec.  The energy density 
per unit wavelength in the observer's frame was reduced by a factor $(1+z)$ 
due to the redshift, representing the ``k-correction'' (Hogg et al. 2002).  
The flux per unit wavelength was then computed according to the inverse square 
of the luminosity distance. Note that we are computing the bandpass energy 
of a single star-particle, not the bolometric magnitude nor its surface 
brightness, which would include an extra factor of $(1+z)$. This is 
essentially  justified by the fact that star-particles are point sources at 
the HST and JWST telescope resolution, as explained in $\S{3.1}$.
While the initial exposure value was set high to extract as much detail as 
possible from the model output, it was later reduced by an appropriate factor 
to take the actual zodiacal background and more realistic exposure times 
into account.  This enabled us to estimate the effect of the sky-background 
on the simulated ``observation.'' For example, for a JWST-class instrument of 
total collecting aperture $A = 25\ \mathrm{m^2}$, the exposure time was 
approximately 11 months for the baseline case when $\emph{no}$ noise was added.
When the standard zodiacal noise for L2 was added, corresponding to H $\approx$ 
22.6 AB-mag arcsec$^{-2}$ (Windhorst et al. 2011), the exposures were reduced 
by a factor of $\approx$ 70 -- 130, depending on the effective width (in Hz) 
of the simulated filter, since the same sky-noise mask with a Poissonnian 
distribution was used to create all the images with sky-background added. 
This corresponds to more realistic exposure times of $\approx$ 2.5 -- 4.5 days 
for JWST deep fields.

In an actual CCD, or near-IR detector, incoming photons are converted to 
electrons, which are then counted. To mimic this, the integrated flux was 
converted to ergs, and divided by $10^{-12}$, approximately the 
energy (`work function'), needed to produce an electron. Since the intent 
was not to simulate exactly any particular instrument, the exact factors  
are unimportant here, provided that the flux and pixel scales are consistent 
with the values used in introducing the sky-background noise, and approximate 
the sensitivity of the simulated instrument class. 
While actual CCDs or IR-detectors have other sources of noise such as 
read-noise and dark current, these are considered second order effects and 
are ignored here. For example, dark current is 10-20 times below zodiacal 
light levels, according to the WFC3 instrument handbook 
\footnote{http://www.stsci.edu/hst/HST$\_$overview/documents}.
While read-noise can be significant for very faint observations, in the 
rest-frame UV we can ignore it for most cases if care is taken regarding the 
length and number of exposures, which would be done in an actual observation, 
also according to the WFC3 instrument handbook for the IR detector. The 
object source also contributes a noise equal to the square root of the source 
signal due to random effects. This can be reduced by very long exposures, 
which we simulate here, so this effect is also ignored, as it is also a second 
order effect. However, this noise is effectively considered when estimating 
the bin count error for the chi-square fit of the simulated LF to a Schechter 
function ($\S{4}$).

This integrated flux gave a simulated `electron count' N$_{e}$, where F is 
the total integrated object flux, and the telescope aperture is given by 
the area A, for an exposure time (t):

\begin{equation}
\  \mathrm{ N_{e} \simeq \ A \ t \ F / 10^{-12} \ ergs}  
\end{equation}
where the incident flux $F_{BP}$
observed through a filter with bandpass BP, 
is given by:

\begin{equation}
\ \ F_{BP} = {\int} F_{\lambda}(\lambda/(1+z),t(z)) R(\lambda) (1+z)^{-1}d\lambda 
\end{equation}
\newline
 \ where $F_{\lambda}(\lambda$) is the incident flux and R($\lambda$) is the filter 
response function. Note that, while the flux calculation is done in the 
wavelength domain, the computation of AB magnitudes from fluxes in 
ergs/cm$^{2}$/sec are done in the frequency domain (Oke \& Gunn 1983).

When trying to exactly simulate an actual
instrument, an optical telescope assembly (OTA) term would need to be
included for the effects of the telescope throughput, and also a 
point-spread-function (PSF) term would need to be included to account
for the mirror figure and any segmentation.
Since no specific instrument is modeled, we simulate the PSF of the WFC3 IR 
camera on Hubble by approximating its PSF by a Gaussian function, which is 
convolved with the FITS image before the sky-noise is added -- as described 
below. Since the WFC3 IR camera has a published FWHM of $\approx$ 0.15" 
(Windhorst, et al. 2011) for the F160W filter, and our simulated image 
pixels are $\simeq$ 0.1", we use a Gaussian with $\sigma$ = 0.65.  While 
the PSF is wavelength dependent and would be smaller for shorter wavelengths, 
this is taken as a worst-case scenario for the lower redshift simulations, 
since we see no significant effect for this filter. This procedure is then 
repeated, but applying the dust-extinction model described in section 3.4. 
The results of both of these are described in section 5.3 and Figs. 4, 5 and 7.

As noted, the space environment is included by adding the proper zodiacal  
sky-background (Windhorst et al. 2011).  Some of the effect of a telescope 
PSF is accounted for in the SExtractor image filter mask, by distributing 
the image over several pixels to increase detection sensitivity to faint, 
low-SB objects. The SExtractor mask used has a FWHM value of 2 pixels, or 
$\approx 0.2"$ for the image scale used here.

\subsection{Simulated ``Observed'' Object Magnitudes}

To compute the ``observed'' apparent magnitude in the AB system, 
the simulated exposure was converted to a flux per unit frequency. 
The integrated flux was divided by the artificial aperture 
and time constants, converting energy units to $\mathrm{ergs\ s^{-1}}$, 
and dividing by the filter effective width in Hz to obtain 
f$_{\nu}$  in $\mathrm{ergs \ s^{-1} cm^{-2} Hz^{-1} }$.

The AB magnitude was then obtained using 
 $\mathrm{m_{AB} = -2.5log(f_{\nu}) - 48.60 }$ (Oke \& Gunn 1983), 
where f$_{\nu}$  is in $\mathrm{ergs \ s^{-1}}$
$\mathrm{cm^{-2} Hz^{-1}}$. To convert to an
absolute M$_{AB}$ magnitude in the rest-frame emitted band, we use 
 $\mathrm{M_{AB}=m_{AB}-5log(D_{L}/Mpc)}$ 
$\mathrm{- 25 + 2.5log(1+z) }$,   
where $\mathrm{2.5 log}$$(1+z)$ is the k-correction, which is positive,
since magnitudes are computed in the frequency domain (Hogg et al. 2002).
In the prior calculation of the total apparent `observed' energy, 
the calculation was in the wavelength domain, since the SEDs 
and filter responses were given as functions of $\lambda$($\AA$).
The calculation of the absolute magnitudes is performed on the photometric 
catalog as produced by SExtractor when detecting `objects' --- clusters of 
star-forming particles --- which are assumed to represent galaxies in formation 
(see discussion below). We used the parameter PHOT\_APERTURES set to 10 
(diameter in pixels) in SExtractor, and the default settings of 
PHOT\_AUTOPARAMS which were 2.5, 3.5. The flux used in the magnitude 
calculation is from the FLUX\_AUTO field output from SExtractor, which is the 
flux within a Kron-like elliptical aperture. Such an aperture is appropriate
for faint largely unresolved objects.

\subsection{SExtractor Parameter and Photometry Testing}

We experimented with several of the SExtractor parameter settings 
by creating catalogs with different settings of either the synthetic FITS
files from the simulation or creating artificial fits images, explained below.
We compared the catalog detection counts and, in some cases, plotted the 
LF from those catalogs to compare the distribution of those detections with
luminosity.

We note that these artificial images represented a different challenge to
detection by SExtractor, compared with real galaxy images. In real images of
galaxies of at least several pixels in size, the light is spread out over
adjacent pixels of the image, until the surface brightness falls below
detection limits. Thus, the object is seen as composed of adjacent or
contiguous pixels, which is required by SExtractor for detection.  However,
our images consist of separate ``star-particles'', where the light is 
concentrated in a single pixel. Without dust and an actual instrument PSF
to diffuse the light into adjacent pixels, smaller objects may be `seen'
by SExtractor to consist of distinct separate pixels. Hence, SExtractor would 
categorize these as a group of stars, or as several smaller source objects, 
rather than a single object. For this reason, Gaussian masks were used to 
diffuse the concentrated sources. This required the testing of different 
size masks.

We tested different Gaussian masks (``FILTER\_NAME'' parameter in the 
SExtractor configuration file) with Gaussian $\sigma$ values ranging from 
1.5 to 5.0 and mask size from $3\times3$ to $9\times9$ pixels. These were 
tested against the FITS image from the simulation at redshift 6.01. The LFs 
were plotted and the results formed two distinct groups. There was negligible
difference in the LF at ABmag brighter than $-14$ mag. Fainter than
$-14$ mag, the Gaussian masks with $\sigma$ values of 1.5 and 2.0 and size of
$3\times3$ pixels showed a sharply reduced number of detections compared
with the other masks, being down an order of magnitude at ABmag $\approx -12$
mag. The mask with $\sigma = 2.0$ and $5\times5$ generally had the most 
detections, hence that mask was used for this study, although the differences 
were small compared with the other masks in that group.

In addition, we ran tests of simulated images with 81 source pixels each 
given a constant flux-count of $10^6$, but spread out over a pattern with  
a variable spacing over a grid of $81\times81$ pixels in a $ 200\times200$
pixel frame. All pixels were separated by at least one, and up to 24 background
pixels. This was then converted to a FITS file for the simulation testing.
This FITS image was scanned with SExtractor using the settings described 
above $\emph{both without}$ adding a PSF-convolution, and $\emph{with}$
a PSF-convolution using a Gaussian with a FWHM of 0.15" and a Poisson 
distributed noise-background with a mean count of $10^4$. Note that the 
source was made constant, not random. This was done to facilitate testing the 
effectiveness and accuracy of the SExtractor photometry, specifically the 
total Kron aperture-flux recovered from the detected objects. That is, we 
could more readily tell if the photometry captured an integer number of source 
pixels and if the background was not being subtracted correctly. By convolving 
with the Gaussian PSF, some of the signal was distributed into neighboring 
pixels. By comparing the SExtractor-calculated photometry of objects for the 
two different detection schemes, we could judge the amount of source flux that 
was not captured by SExtractor.  It was found that before being convolved with 
the PSF, SExtractor found 10 objects with a total flux-count of 
${4.0\times 10^7}$. Converting flux units to magnitudes, SExtractor found an 
average difference of ${0.00004 \pm 0.00009}$ mag compared with what would be 
expected if an integer number of sources were found, before convolving with 
the PSF. After convolving with the PSF, SExtractor found 29 objects with a 
total flux count of ${9.37\times10^7}$, and an average difference of 
$0.003 \pm 0.002$ mag compared with the expected value. This test was repeated 
with higher background noise levels. With a mean background count of $10^5$, 
before convolving with the PSF, SExtractor found only 5 objects, reflecting 
the lower S/N, and an average difference of $0.003 \pm 0.02$ mag compared with 
the expected value.  After convolving with the PSF, SExtractor found 12 objects
and an average difference of $0.02 \pm 0.01$ mag compared with the expected 
value.  When a background with a mean count of $10^6$ was used, SExtractor 
found 12 objects with an average difference $-0.02 \pm 0.01$ mag compared with 
the expected value. 

Finally, we simulated objects closer to the S/N range of the fainter
images created in the numerical simulation. A fixed flux was still used, 
in order to test the SExtractor photometry, but with an integrated flux count 
of only $10^4$, which is in the range of the simulated star-particle
fluxes. For example, the simulation at redshift 6.24, before any adjustment to
the simulated exposure time, with the simulated WFC3 IR filter, had fluxes 
ranging from $3.5\times10^2$ to $3.1\times10^6$ with an average flux count of 
$8.9 \times 10^4$. An object detected by SExtractor in the numerical 
simulation with a flux count of $1.0\times10^6$ was at a simulated apparent 
AB magnitude of 32.76 mag and an absolute magnitude of $-14.02$ AB-mag.  
After the simulated source signal was reduced by a factor of 83 for the 
expected background noise for that filter, and to simulate a more realistic 
exposure time, as discussed in $\S{3.2}$, a source count of $1.0\times10^6$ 
corresponded to m$_{AB}$ of 27.60 mag and M$_{AB} = -19.18$ mag. These data 
are given to provide a comparison of the artificial test images with the the 
simulated images.

A much larger number of particles was then simulated, with variable spacing 
but more densely packed. Also, a random floating point value in the range from 
$-1.0$ to $+1.0$ pixels was added to the X and Y coordinates to each source 
pixel. A random background with a mean count of $10^6$ was added and the image 
was also convolved --- before adding noise --- with the PSF for comparison.  
Without the PSF, SExtractor found 18 objects with an average flux count of 
$0.72\times10^6$  and a difference of $-0.002 \pm 0.002$ mag compared with 
the expected value. After the PSF-convolution, SExtractor found 8 objects 
with an average flux count of $0.80\times10^6$ and a difference of 
$-0.002 \pm 0.002$ mag compared with the expected value.

Therefore, we conclude that SExtractor photometry on these types of objects
consisting of discrete relatively high S/N sources, shows no significant flux
difference, even after convolving with a PSF. When the S/N ratio is reduced
with the smaller samples we see a systematic error, but this may be due to the 
smaller sample size, since we again see a small difference with the larger 
samples even with the higher background--level and smaller source pixel flux 
count. In addition, we found best results for the parameter DETECT\_THRESH of 
1.5 sigmas above background for pixel detection, which was the default 
setting. We also found a DEBLEND\_NTHRESH setting of 64, rather than the 
default setting of 32, slightly increased the number of detections.

\subsection{Processing of the Simulated ``Observed'' Images}

The files of flux and sky-projected object coordinates were translated into 
FITS files using publicly available utilities from the High Energy 
Astrophysics Science Archive Research Center of NASA 
\footnote{http://heasarc.nasa.gov/lheasoft}. 
Their size was set by the number of pixels along the sky--projected axes, 
according to the comoving distance at that redshift and the $\Lambda$CDM model 
used, as described above. The FITS object frame or merged frames were combined 
with a simulated sky-noise background FITS file. This noise frame was computed 
using a Poisson distribution.  As explained previously, a convolution mask was 
used in SExtractor to improve the classification of close groupings of
star--particles, including normal galaxies and mergers, as extended objects 
rather than as isolated star--like objects.
This was 
necessary, since gas particles and radiative transfer are not currently 
incorporated in the simulated images. 

Various parameters were used to extract the simulated objects, and inspect the 
sensitivity of the results to the input parameter selection. This is a 
difficult problem, since we have two main independent sources of error: 
the simulation itself and the selection criteria used for SExtractor. 
A gaussian convolution mask of $5 \times 5$ pixels with a full width half 
maximum (FWHM) of 2.0 pixels was used. This enables star-particles which are 
near each other, but not {\textquotedblleft}touching{\textquotedblright}, to be
detected as part of a ``single object'' (see Fig. 1). We also set the SExtractor
object detection at a minimum of 40 pixels above threshold (DETECT\_MINAREA). 
This was done to remove any spurious identification of pixel groupings
which were actually part of the same object as a collection of smaller
objects.
At the chosen pixel scale of 0.099"/pix, our 2.0 pixel or $\approx$ 0.2" 
convolution kernel, is larger than both the PSFs of HST (0.13" FWHM) or JWST 
($\approx$ 0.06" FWHM) at $\approx 1\mu$m wavelength, which corresponds to 
$\approx$ 2.2 kpc convolution kernel, i.e., larger than most SF regions of 
interest. It is similar to the sizes of faint galaxies expected at these 
magnitudes (e.g., Windhorst et al. 2008).

\subsection{Dust Extinction}

The effect of dust extinction on the LF is modeled by using the relation 
from Calzetti et al. (2000) between the UV extinction (A1600) and
the slope ${\beta}$ of the UV continuum, and by using the study of the
UV continuum ${\beta}$ in Finkelstein et al. (2012).

The data from Finkelstein et al. (2012) is modeled both in terms of 
redshift and luminosity, with interpolations made between their redshift 
bins, to yield a continuous function across the luminosity range.  
The extinction is calculated using the median values for ${\beta}$
in Finkelstein et al. (2012), and also using a Monte Carlo selection
of ${\beta}$ over the calculated scatter in their values.
The scatter or spread in the Finkelstein et al. (2012) ${\beta}$ values is
approximated by a normal distribution with a mean value equal to the 
median value they give, and a standard deviation equal to the square root 
of the average of the squares of their positive and negative one-sigma 
values. These are typically $\sigma_{\beta} \approx 0.2$.

Since Finkelstein et al. (2012) caution that their ${\beta}$--values at 
$z \ge 8$ are unreliable, we do not use those, and use the results at 
$z =$ 7.0 for redshifts $z$ = 7.16 and 7.68.  As they consider dust extinction 
rather unimportant for the higher redshifts, we do not apply a dust 
extinction to the simulated objects at $z$ = 10.38.
The ${\beta}$ values in Finkelstein et al. (2012) ranged from --1.8 at 
redshift 4 to --2.68 at redshift 7. Our simulated A$_{UV}$ values range
from 0.64 mag at redshift 4.52 to 0.006 mag at redshift 7.68, and 0 at 
redshift 10.4. See Table 3 for a breakdown of A$_{UV}$ values and 
their 1--$\sigma$ errors, by magnitude range and type of simulation.

The results are given in Sections 5.3 and in 5.4 with a PSF added, and do 
not show a significant difference in the derived faint--end LF slope alpha--
values, even though Table 3 shows differences in the A$_{UV}$ extinction by 
magnitude. However, there appears to be an effect from dust on the derived 
$M^{*}$ values at the lower ($z$ = 4.5 and 5.3) redshifts.

We note that some authors (e.g., Jaacks et al. 2012 and Shimizu et al. 
2014) adjust the dust extinction values within a given parameter range
to obtain the best fit of their simulated LFs with the observed LFs. We
take a different approach in order to preserve the predictive nature
of our model. This results in number densities per unit volume that
are significantly higher than in the literature. If we adjusted our
dust extinction per the prescription in Jaacks et al. (2012) based on 
a range of E(B--V) values of 0.0 to 0.30, we would have increased our
A$_{UV}$ to 2.33 mag, and come much closer to predicting the observed LFs.  
They ended up using E(B--V) = 0.10 for redshifts 6 and 7, which resulted in 
A$_{UV}$ = 0.23 mag, but their choice was based on that value resulting in 
a better fit of their simulated LFs to observed LFs.  We also note that our 
approach results in a luminosity--dependent extinction function, rather than an 
overall extinction, so that it could affect the Schechter fit of the LF, but,
as discussed later in $\S{5.3}$, we do not observe any significant trend here.

\begin{table*}[t]
\begin{center}
\caption{Simulated A$_{UV}$ corrections.}
\begin{tabular}{rcccr}\hline
\multicolumn{5}{c} {} \\ %
Redshift & $M_{AB} \le-20$ & $-20<M_{AB} \le -18$ & $-18<M_{AB}$ & Comments \\\hline 

4.52 & $0.477 \pm 0.131$ & $0.612 \pm 0.155$ & $0.640 \pm 0.158$ & no winds\\\hline
5.32 & $0.352 \pm 0.183$ & $0.260 \pm 0.214$ & $0.251 \pm 0.234$ & no winds\\\hline
6.01 & $0.201 \pm 0.209$ & $0.077 \pm 0.144$ & $0.021 \pm 0.087$ & no winds\\\hline
6.01 & $0.194 \pm 0.209$ & $0.077 \pm 0.140$  & $0.023 \pm 0.087$ & winds\\\hline
6.24 & $0.156 \pm 0.208$ & $0.056 \pm 0.134$  & $0.019 \pm 0.090$ & no winds \\\hline
6.24 & $0.150 \pm 0.207$ & $0.058 \pm 0.136$  & $0.018 \pm 0.085$ &  winds \\\hline
7.16 & $0.121 \pm 0.208$ & $0.021 \pm 0.093$ & $0.014 \pm 0.085$ & no winds \\\hline
7.16 & $0.112 \pm 0.198$ & $0.022 \pm 0.098$ & $0.010 \pm 0.074$ & winds \\\hline
7.68 & $0.136 \pm 0.234$ & $0.019 \pm 0.092$ &  $0.012 \pm 0.073$ & no winds \\\hline
7.68 & $0.123 \pm 0.216$ & $0.017 \pm 0.085$ &  $0.006 \pm 0.056$ & winds \\\hline
10.38 & 0 & 0 & 0 & \\\hline
\end{tabular}
\newline
\newline
Simulated extinction values A$_{UV}$ using UV continuum slope $\beta$ values 
from Finkelstein 
et al. (2012). 
\newline
Comments column indicates whether simulation includes feedback winds 
or does not. 
\newline
For details, see text ($\S{3.6}$).
\end{center}
\end{table*}

%
\begin{figure*}[h]
\epsscale{.50}
\plotone{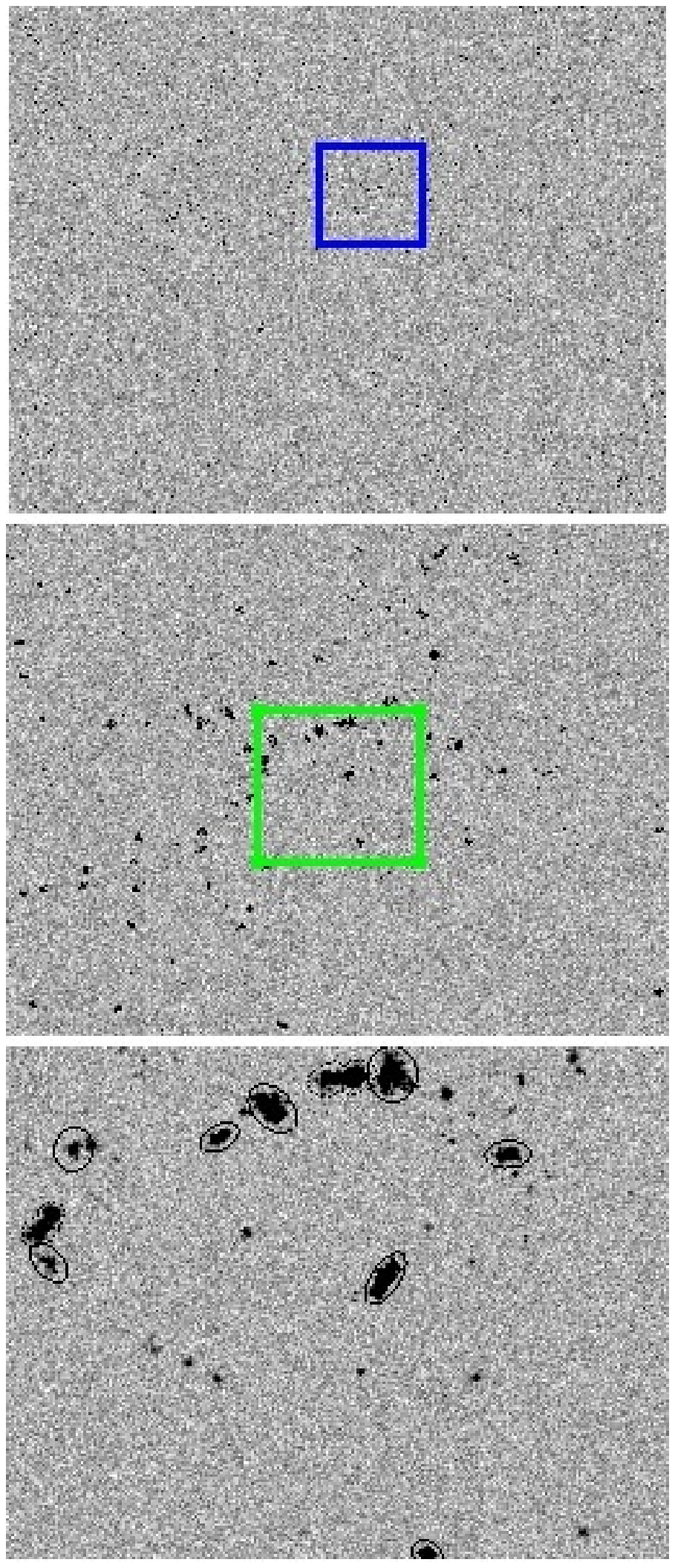}
\caption{FITS file and image aperture file output from SExtractor
          at redshift ${ z = 6.01}$ through the simulated WFC3 F105W 
	  filter with adopted sky-background 
	  $\mathrm{\approx 22.6 \ m_{AB} \ arcsec^{-2} }$ 
	  added to the original simulated image.
	  The image on the top is the entire simulated field 
	  $ \approx $  110 arcmin$^{2}$ in size.
	  The portion in the blue box is enlarged in the middle image, 
	  $\approx$ 78 by 90 arcsecs. The portion in the green box is shown 
	  on the bottom, $\approx$ 24.5 by 23.4 arcsecs. Dark pixels are 
	  individual `star particles', treated as SSPs (see $\S 2.1$).  
	  Notice that the images in the contours have luminosity concentrated 
	  in the simulated `star particle' pixels and lack the extended surface
	  brightness ``wings'' of actual observed images. }
\end{figure*}

\section{Schechter LF Fitting to the Derived ``Observed'' LFs}

We fit the LF to a Schechter function by minimizing the chi-square
value of the sum of the squares of the differences between predicted counts
and object counts in magnitude bins, divided by the predicted bin count. 
A variable bin-size was used to maximize the degrees of freedom,
and to ensure a minimum count per bin to improve the reliability
of the fit. A maximum bin count was later imposed, to improve the fit stability
by reducing the variability of the bin-counts due to `bunching' of data around 
the bin magnitude limits, as discussed in more detail below. We also performed 
a best-fit, selected over the ``observed'' data parameter space, which is also 
discussed below.

The photometric data collected from the SExtractor catalogs were converted 
to restframe absolute magnitudes.  The filters were chosen such that the 
sampled restframe emission band was approximately 150 nm, in order to allow 
comparison with the Hubble WFC3 ERS data of Hathi et al. (2010), and references
therein. In Hathi et al. (2010), the authors explain that they used the dual 
image mode of SExtractor, where one set of images was used for detection and
another for the photometry. This was due to the use of multiple filters and 
the use of MULTIDRIZZLE.  They used multiple filters to use the drop-out 
technique to find candidates for different redshift bins. We already know the 
redshift of the object a priori, so we mimic what they did, by using the single
image mode, in a single filter for each redshift that most closely samples 
restframe 150 nm.

We fit the LF data to a Schechter function in magnitude space of the form:
\begin{eqnarray}
\Phi (M) & = &  0.4 \ln(10) \ \phi^{ \ast} \exp \ (-10^{[-0.4 (M - M^{\ast})]})  \times \nonumber \\ &  & 10 \ ^{[(-0.4) (\alpha + 1)(M - M^\ast)]}  
\end{eqnarray}
where $M$ is the absolute magnitude, and $M^{\ast}$ is the characteristic 
magnitude of the Schechter LF.  $\Phi(M)$ is the volume density count of 
objects of magnitude $M$, with $M^\ast$, $\alpha$ and the normalization 
$\phi^{*}$ as free parameters. The fit used the chi-square function:
\begin{equation}
	\chi^{2}  =  \Sigma [Y_{i} -  y_{i}(\theta) ]^{2}/y_{i}(\theta)  ,
\end{equation}
where $Y_{i}$ are the measured values, and $y_{i}(\theta)$ are the expected  
values for parameters $\theta$, which here are $M^{\ast}$, $\alpha$ and 
$\phi^{*}$, respectively. 
The expectation function is the Schechter (1976) function.
The free parameters are the characteristic magnitude $M^{\ast}$, where the 
exponential function `breaks', the slope $\alpha$ of the faint-end power-law, 
and the normalization value $\phi^{*}$.

The minimization technique was essentially a brute force calculation over 
a broad magnitude range of the parameter space 
(${-33.9 \leq M^{\ast}_{AB} \leq -16.9}$ mag with 500 equal steps and 
${-2.4 \leq \alpha \leq -1.0}$ with 150 steps),
calculating the sum of the residuals for each parameter combination. 
For simulations where the error was very large and $\alpha$ was very steep, the 
range of $\alpha$ was extended to $ -3.0$ to avoid biasing the error.
For each pair of $M^{^\ast}$ and $\alpha$, a dynamic fitting was used to 
minimize the chi-square for the normalization parameter $\phi^{*}$. 
The code `zoomed in' (i.e., took smaller steps)
when the chi-square value fell below a specified threshold value, and then 
exited when the chi-square value exceeded another threshold after reaching 
a minimum. This method was checked against a non-dynamic, but much slower 
search. Chi-square contours were drawn for three confidence levels 
(68\%, 90\%, and 99\%)  by adding the appropriate increment (3.50, 6.25, 
and 11.30, respectively) to the best-fit minimum chi-square value (see e.g., 
Practical Statistics for Astronomers, Wall \& Jenkins 2012). These are the 
$\mathrm{\chi ^2_3(significance) }$ values  for the 3-dimensional parameter 
space $M^{*}$--$\alpha$--$\phi^*$ for $\mathrm{significance = 0.32, 0.10, \ and 
\ 0.01} $, respectively, although we plot only the projection in the 
2-dimensional parameter subspace $M^{*} - \alpha$.

The fits were also performed by varying parts of the simulated parameter space, 
namely the maximum absolute magnitude. This was necessary, since the LF 
drops off steeply at faint magnitudes, generally around mag AB $\simeq$ --16.0 
to --17.0 with minimal sky-noise (${\approx 40}$ ABmag) and at 
$\mathrm{AB \approx -18.0}$  for a more realistic sky-background in space 
($\mathrm{\approx \ 22.6 AB-mag/arcsec{^{2}}}$). The cutoff magnitude was found 
by fitting a Schechter function to the data, and repeating the process, 
making the cutoff magnitude on the faint--end about 0.25 magnitudes fainter.  
The cutoff was found by finding the maximum magnitude range for which the 
reduced chi-square fit was equal to, or less than, a factor of $\sim$ 2.0 for 
all the LF-redshifts for a given test category (i.e., basic simulation, 
sky-background added, and $\emph{with}$ feedback winds). In most cases, the 
reduced chi-square was 1.5 or less. 

This dropoff or turnover at the faint--end of the ``observed'' LF is apparently 
partly due to incompleteness from a combination of the model pixel resolution 
and the effect of the imposed sky-background noise.  We note, however, that 
a higher resolution simulation ($\mathrm{\approx 10^3\ M_{\odot}}$ mass per 
particle) also including gas cooling, heating and star-formation by Read 
et al. (2006) found the smallest building blocks to occur at 
$\mathrm{\approx 10^8\ M_{\odot}}$ with a stellar mass of 
$\mathrm{\approx 10^6\ M_{\odot}}$, which is very close to our effective
mass limit. The actual sky-background-induced cutoff was $\approx$5---6 
magnitudes brighter than the minimum magnitude of the ``detected'' simulated 
objects, which were ``detected'' to $\approx$ --10 to --12 AB-mag with the 
case of minimal noise-added.  This would seem to indicate a preference in the 
model for forming `medium' sized objects, rather than just a continuation of 
the power law at the faint-end. This is discussed in more detail later 
($\S{6}$). 

Each bin's effective absolute magnitude was given by the average over the 
absolute magnitudes for each ``detected'' object in that bin. To increase 
the reliability of the chi-square fit, a minimum of 5 objects were required 
for each bin with a minimum bin size of 0.1 magnitude.  To improve the 
stability of the chi-square fit over the faint-end magnitude range, a variable 
sized magnitude bin was used.  This was achieved by imposing a maximum limit on 
the number of objects per bin, which reduced the variability of the chi-square 
value, and improved the consistency of the cut-off magnitude for the best-fits 
for the different redshift samples. Generally, about ${1 \%}$ of the total 
number of objects detected by SExtractor was used as the maximum bin-count 
limit, varying from 60 to a minimum of 15. Trials were used with a range of 
maximum counts to determine the most stable fit,  resulting in approximately 
twice the number of bins as when the mimimum bin size of 0.1 magnitude was 
used. Thus, the degrees of freedom were also increased, contributing to 
better chi-square fits in this case.

The uncertainty for each parameter $M^{\ast}$ and $\alpha$ was found by 
projecting the  1--sigma contour orthogonally onto each parameter axis. In 
practice, this amounted to a search in the $M^{\ast}$ -- $\alpha$ parameter 
space for chi-square values bracketing the one sigma (68\%) values described 
above. Chi-square values computed in the search were captured in an array of 
minimal chi-square values for each $M^{\ast}$, $\alpha$ pair. The minimum was 
found by searching over $\phi^{*}$ -- the LF normalization factor. 
The LFs and best-fit Schechter functions are shown in figures 2 --- 7, both 
\emph{with} and \emph{without} added zodiacal background noise, and also, 
\emph{with} additional feedback physics. Confidence contours for the 
Schechter function parameters $M^*$ and $\alpha$ for the fits are also shown.

\section{LF Results: Simulated Faint-End LF Slope Evolution}

We examine the results of the numerical simulation: 
\begin{itemize}
\item \emph{without} feedback in the form of winds (\S{5.1}), and 

\item  \emph{with} feedback (\S{5.3}). 

\item  with adding simulated sky noise from the zodiacal background 
to the mock images created from the output of the simulations (\S{5.2 \& 5.4}).  
\end{itemize}

LFs were fitted to the 
``observed'' images from the simulations with restframe emission at 
150 nm (UV) at redshifts in the range of $z \simeq$ 4.5 to 10.4, as shown 
in Figs. 2 --- 7. In Fig. 8 we compare our simulated LFs with more recent 
observed LFs from Finkelstein et al. (2014) at redshifts 6 to 8. In 
figures 9 --- 11 we show the evolution of the best-fit faint-end LF-slope 
$\alpha$ as a function of redshift and compare with the observations 
(Hathi et al. 2010) and references therein. For a summary of the results,
see Table 4.

\subsection{Simulated Faint-End LF Slope -- No Feedback, No Sky-Noise}  
We first look at the LF fits to the simulation results without feedback `winds'
and with extremely low sky-background (Fig. 2), where a minimal amount of 
sky-noise is added only for SExtractor functionality (e.g., Tamura et al. 2009).
In Fig. 9, we compare the redshift evolution of $\alpha$ from our `no-winds' 
model results to the observations from Hathi et al. (2010).  We find very 
similar results to the observed Hathi et al. (2010) fits for the faint-end 
slope $\alpha$ of the LFs, who found a faint-end slope in the range of 
$\alpha \approx $ --1.2 to --1.8, with a redshift dependence of: 
\begin{equation}
  \mathrm{  \alpha = -1.10 - 0.10 \ z \  \     [observed]}
\end{equation}
Our results for the initial `no-winds' model over the redshift range
of \emph{$z \simeq$ 4.5 to 7.7}, are (Fig. 9): 

\begin{eqnarray}
\mathrm{\alpha  =  -1.00 \pm 0.14 - (0.13 \pm 0.02) \ z} \\   \mathrm{ [Simulated - No\ Winds]} \nonumber
\end{eqnarray}
There appears to be some evidence of a lessening of the $\alpha$--redshift
dependence at higher redshifts, as seen in Figs. 10 \& 11.
For redshifts $\mathit{ 6.0 \le z \le 10.4 }$, we find:
\begin{eqnarray}
\mathrm{\alpha  = -1.44 \pm 0.19 - (0.06 \pm 0.03) \ z}  \\    \mathrm{[Simulated - No \ Winds]} \nonumber
\end{eqnarray}
This is discussed in more detail later, when we discuss 
the simulations that include feedback in the form of `winds' in the model.

The derived or implied characteristic magnitude, $M^{\ast}$, does not 
appear to correspond well with the actual observations (Hathi et al. 2010, 
Bouwens et al. 2011, 2014). However, we note that this value
is generally outside of --- or near the boundary of --- the bright-end of the 
magnitude range of the simulated data, as seen in Fig. 2, and also
Figs. 3 --- 7. Therefore, $M^{\ast}$ may just be an indication of where the 
bright-end LF ends due to the lack of dynamic range in the simulated data, 
caused by the limited simulation volume as necessitated by the available 
computer resources.

Thus, it is likely that $M^{\ast}$ and its evolution is not reliably treated 
by the simulated data thus far, and that any apparent dependence of $M^{\ast}$ 
on redshift is mainly due to the changes in the available simulated magnitude 
interval with redshift. Also, we note that there are increased numbers of 
detected objects at brighter magnitudes as the redshift decreases, likely due 
to merger activity developing larger, and hence generally brighter, objects.

We see that the number densities per volume are an order of magnitude or
more greater than observed (e.g., Hathi et al. 2010, Bouwens et al. 2011).
In following sections, we apply dust extinction and feedback in the form of 
``winds'' to attempt to reduce this overproduction.  This overproduction 
is a common problem in numerical simulations without feedback, usually
characterized as an ``overcooling'' problem (SH03).

\begin{table*}[ht]
\begin{center}
\caption{Simulated faint--end LF slope $\alpha$(z).}
\begin{tabular}{rcc}\hline
\multicolumn{3}{c} {} \\ %
Model Description & $ 4.5 \le z \le 7.7 $ & $6.0 \le z \le 10.4$ \\\hline 
No Sky--BG, PSF & ${-1.00 \pm 0.14 - (0.13 \pm 0.02) \ z}$ & 
${-1.44 \pm 0.19 - (0.06 \pm 0.03)\ z}$ \\
or Winds &  & \\\hline
With Sky--BG & ${-1.17 \pm 0.20 - (0.10 \pm 0.03) \ z}$ & 
${-1.56 \pm 0.28 - (0.04 \pm 0.04)\ z}$ \\
No PSF, Dust, or Winds & & \\\hline
With Sky--BG \& Dust  & ${-1.15 \pm 0.23 - (0.11 \pm 0.03) \ z}$ & 
${-1.80 \pm 0.28 - (0.02 \pm 0.04)\ z}$ \\
No PSF or Winds & & \\ \hline 
With Sky--BG \& Dust  & ${-1.16 \pm 0.22 - (0.12 \pm 0.03) \ z}$ & 
${-1.79 \pm 0.27 - (0.03 \pm 0.04)\ z}$ \\
\& PSF, No Winds & & \\ \hline \hline
With Winds, No PSF,  & N/A & ${-1.54 \pm 0.25 - (0.06 \pm 0.04) \ z}$  \\
Sky--BG, or Dust & & \\\hline
With Winds, Sky--BG,  & N/A & ${-1.84 \pm 0.44 - (0.04 \pm 0.07) \ z}$  \\
PSF \& Dust & & \\\hline
\end{tabular}
\newline
\newline
The faint-end LF slope $\alpha$(z) from best-fit Schechter functions
fitted to the model outputs from the basic no feedback winds model
and the model including feedback winds. Variations are also shown for adding
zodiacal sky-background, a simulated PSF of $\approx$ 0.15'' FWHM and simulated 
dust extinction (Table 3). Note the agreement in $\alpha$(z) evolution -- 
the coefficient of z -- in the redshift range $4.5\le z \le 7.7$,
and the closeness to the observed $\alpha$(z) evolution from Hathi et al. 
(2010), who found $\alpha = -1.10 - 0.10 \ $z.  Note, also, the agreement, 
within 1-$\sigma$, in the z-coefficient over the redshift range 
$6.0\le z \le 10.4$, and that it suggests a shallower evolution than the lower 
redshift range. For details, see $\S{5}$.
\end{center}
\end{table*}

\subsection{The Case of Sky-Background - Without `Winds'}

In order to compare the model results with more realistic space-based 
observations, it is necessary to introduce sky-noise (the Zodiacal 
sky-background) into the images.  This is shown in Fig. 3.
As previously discussed, this was done by adding a randomly generated noise 
frame to the previously generated FITS images from the `no winds' simulation, 
by carefully adjusting the data-signal levels to achieve the appropriate 
background level of $\mathrm{\approx 22.6 \ AB}$-$\mathrm{mag \ arcsec^{-2}} $ 
at the near-IR `detection' wavelength of 1.6 $\mu m$ (H-band). While this has 
the expected effect of truncating the LF at about M$_{AB} \simeq -18$ mag,  
we now obtain over the redshift range $\mathit{ 4.5 \le z \le 7.7 }$ (Fig. 10):
\begin{eqnarray}
\mathrm{\alpha = -1.17 \pm 0.20 - (0.10 \pm 0.03) \ z} \ \ \ \ \ \  \\   
\textrm{ [Simulated -- No Winds, With Sky background]} \nonumber
\end{eqnarray}
which is slightly steeper than the results observed by Hathi et al. (2010),
but is in agreement with respect to the evolution of $\alpha$(z).

We note that the 1--$\sigma$ error at $z = 10.38$ has increased substantially 
over the \emph{no sky background} case, apparently due to incompleteness 
effects induced by the added sky-background.  This is also reflected in the 
error contour maps for $\alpha - M^*$, which may indicate that more 
sophisticated fitting methods than chi-square minimization might be needed.

This result compares with the best-fit faint-end slope $\alpha$ over the higher
redshift range of $6.0 < z < 10.4$:
\begin{eqnarray}
\mathrm{ \alpha (z) = -1.56 \pm 0.28 - ( 0.04 \pm 0.04) \ z } \ \ \ \ \ \  \\ 
\textrm{[Simulated -- No Winds -- With Sky background]} \nonumber 
\end{eqnarray}
We observe that the evolution of $\alpha$(z) with redshift has become flatter 
in the higher redshift range in our simulated data when compared to Eq.(8).

We note that $M^*$ is essentially unaffected by the addition of sky-background
--- except that, as noted above --- the error in the fit is increased 
due to faint-end incompleteness effects induced by the sky-background. 
Comparing the counts of detected objects in the samples at different redshifts
\emph{with} and \emph{without} sky-background added enables us to estimate 
completeness of the sky-background samples (Table 2).
Since the simulations essentially sample star-particles of 
${2.7 \times 10^{5} h^{-1} M_{\odot}}$ or M $\leq -12$ mag (see Fig. 2), while 
the actual data (sky background) limits the LF to be sampled to M $\leq -18$
mag (see Figs. 3 --- 7), the ratio of the two (Table 2) helps us to assess the 
incompleteness of the ``real world'' samples at M $\geq -18$ mag.

\subsection{The Case of Dust Extinction and PSF - Without `Winds'}

The effects of dust extinction, described in $\S{3.5}$, were added to the 
catalogs of objects selected by SExtractor from the simulated images with
sky-background added (Fig. 4). Later, we also show the effects of adding a 
simulated PSF to the synthesized images (Fig. 5). The resulting LFs were fitted
to a Schechter function and best-fit parameters derived as before, with a 
limiting absolute AB magnitude M$_{AB}\approx$ --18.5.  
We obtain over the redshift range $\mathit{ 4.5 \le z \le 7.7 }$ (Fig. 10):
\begin{eqnarray}
\mathrm{\alpha = -1.15 \pm 0.23 - (0.11 \pm 0.03) \ z} \ \ \ \ \ \ \ \ & &    \\ 
\textrm{ [Simulated -- No Winds, With Sky-BG and Dust]} &  & \nonumber
\end{eqnarray}
We observe that this result is very similar to Eq.(8), indicating that the
addition of dust extinction has not significantly affected our results.
This result compares with the best-fit faint-end slope $\alpha$ over the 
higher redshift range of $\mathit{6.0 \le z \le 10.4}$:
\begin{eqnarray}
\mathrm{\alpha = -1.80 \pm 0.28 - (0.02 \pm 0.04) \ z}  \ \ \ \ \ \ \ \ \ &  & \\ \textrm{ [Simulated -- No Winds, With Sky-BG and Dust]} \nonumber
\end{eqnarray}

When a simulated PSF of $\approx$ 0.15" FWHM is added to the images --- before 
adding sky-background --- and then the resulting catalog has dust extinction 
applied, we obtain --- over the redshift range $\mathit{ 4.5 \le z \le 7.7 }$:
\begin{eqnarray}
\mathrm{\alpha = -1.16 \pm 0.22 - (0.12 \pm 0.03) \ z} \ \ \ \ \ \ \ \ & &  \\
\textrm{ [Simulated -- No Winds, With Sky-BG, Dust and PSF]} \nonumber
\end{eqnarray}
Again, we note the similarity of this result to Eqs.(8) and (10), indicating
that the addition of the simulated PSF makes little difference to our results,
likely due to the nature of our simulated images and the use of the Gaussian
filter in SExtractor, as discussed previously.

This result compares with the best-fit faint-end slope $\alpha$ over the 
higher redshift range of $\mathit{6.0 \le z \le 10.4}$:
\begin{eqnarray}
\mathrm{\alpha = -1.79 \pm 0.27 - (0.03 \pm 0.04) \ z} \ \ \ \ \ \ \\
\textrm{[Simulated -- No Winds, With Sky-BG, Dust and PSF]} \nonumber
\end{eqnarray}
We note that these compare very closely to the prior results without UV dust
extinction and without the PSF applied to the images before SExtractor. This 
is because the median faint galaxy size is somewhat larger than the PSF\_FWHM,
so that convolving with the PSF does not make a huge difference, other than 
perhaps a slight loss in sensitivity.

\begin{figure}
\begin{center}
\epsscale{0.98}
\plotone {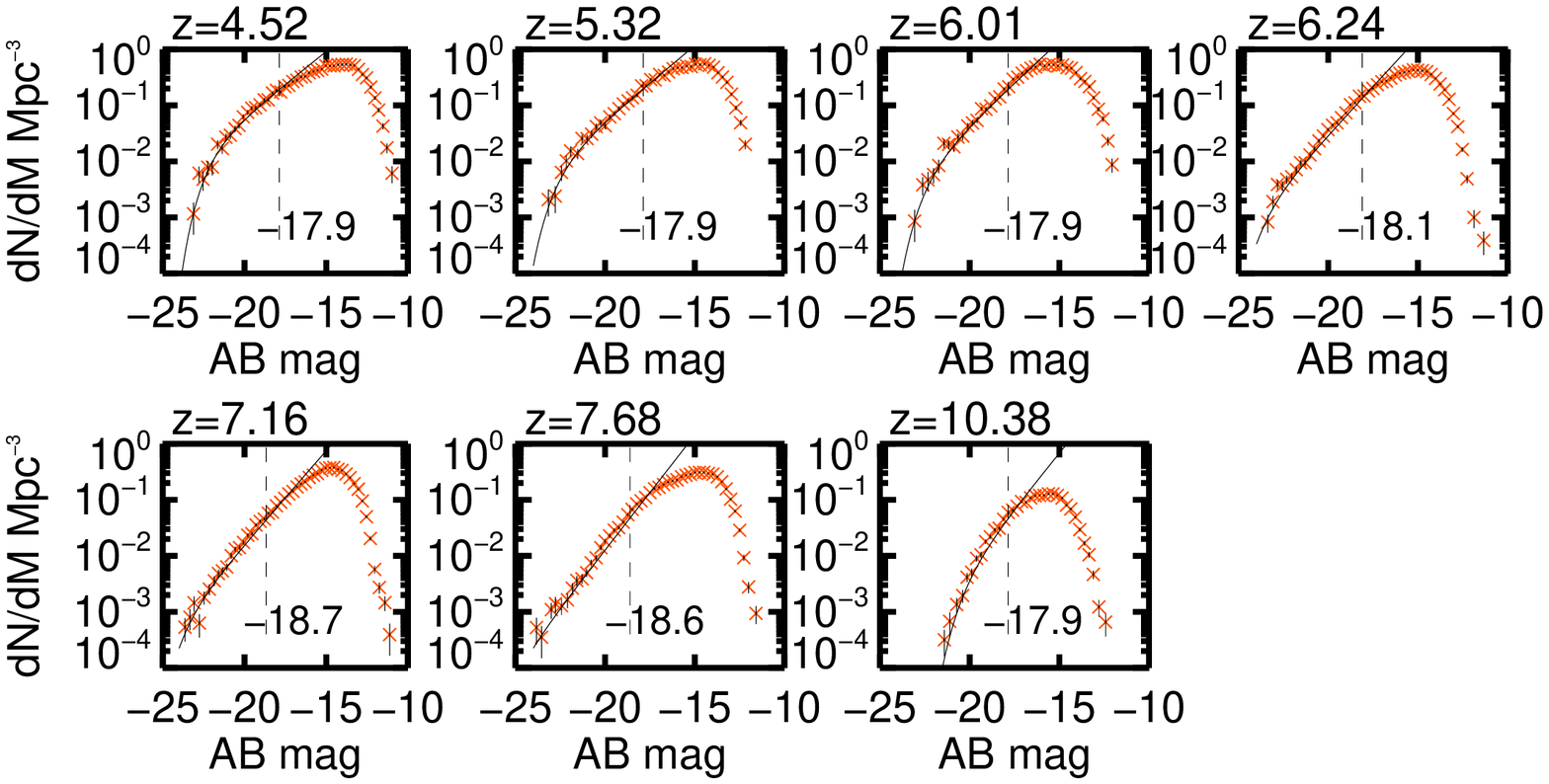}
\epsscale{0.94}
\plotone {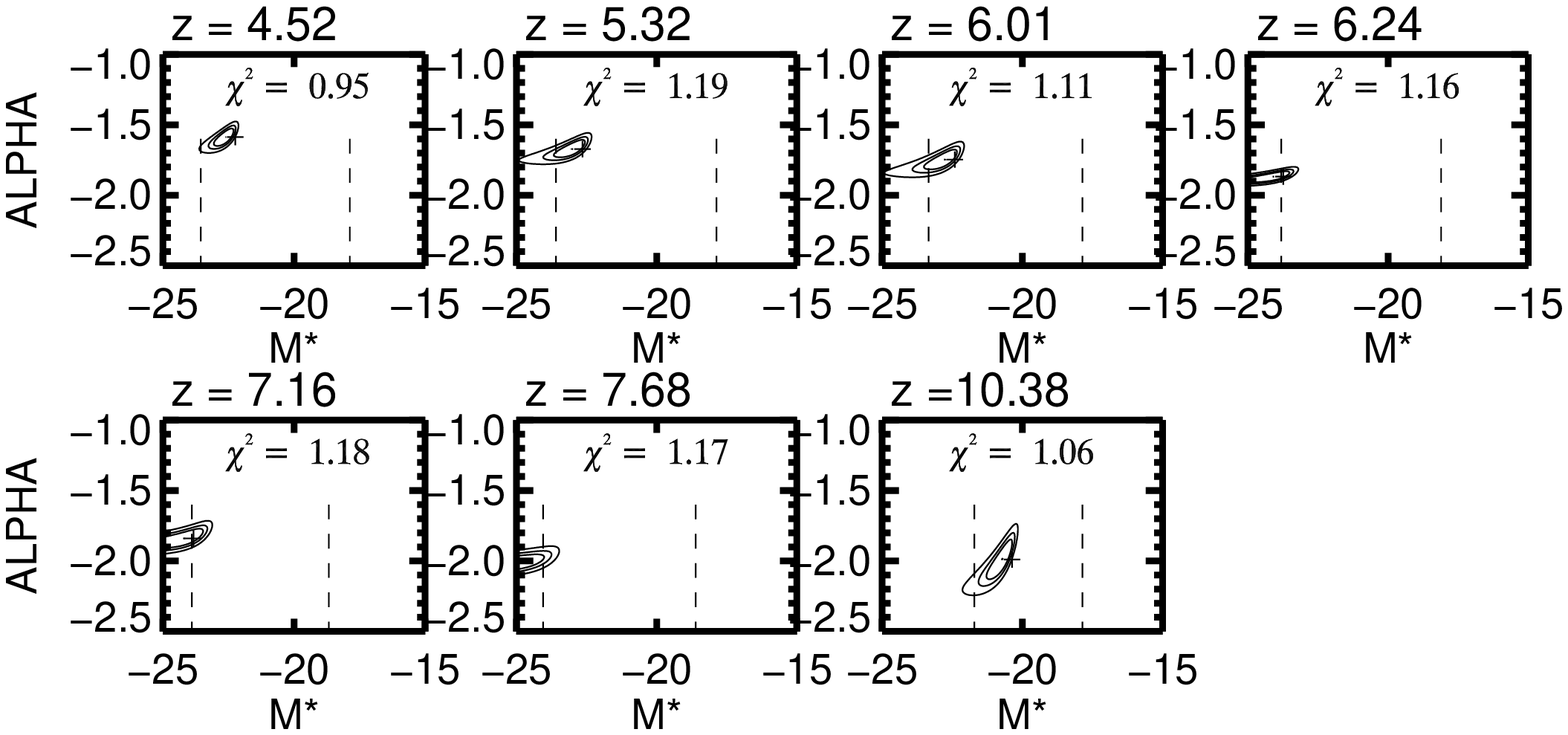}
\caption{
The Luminosity Function ${dN/dM/Mpc^3}$ vs. AB-mag of simulations 
\emph{without feedback winds} and the confidence-region panels of the best-fit 
Schechter functions in the $\alpha$--${M^*}$ parameter space. The contours 
indicate 68\%, 90\%, and 99\% confidence levels. The best-fit Schechter 
functions are given by the solid black curves in the upper panels. 
These are from images created \emph{without} adding sky-background noise.
The LFs are at numerically simulated redshifts of 4.5, 5.3, 6.01, 6.24, 7.16, 
7.68, and 10.38. The vertical dotted lines in the upper LF plots indicate the 
faint-magnitude cutoffs used, when obtaining a chi-square fit to the Schechter 
function with a reduced chi-square value of no more than 1.2, and an absolute 
magnitude brighter than $\mathrm{M_{AB} \simeq -17.5 mag}$.}
\end{center}
\end{figure}
\begin{figure}
\begin{center}
\epsscale{0.98}
\plotone {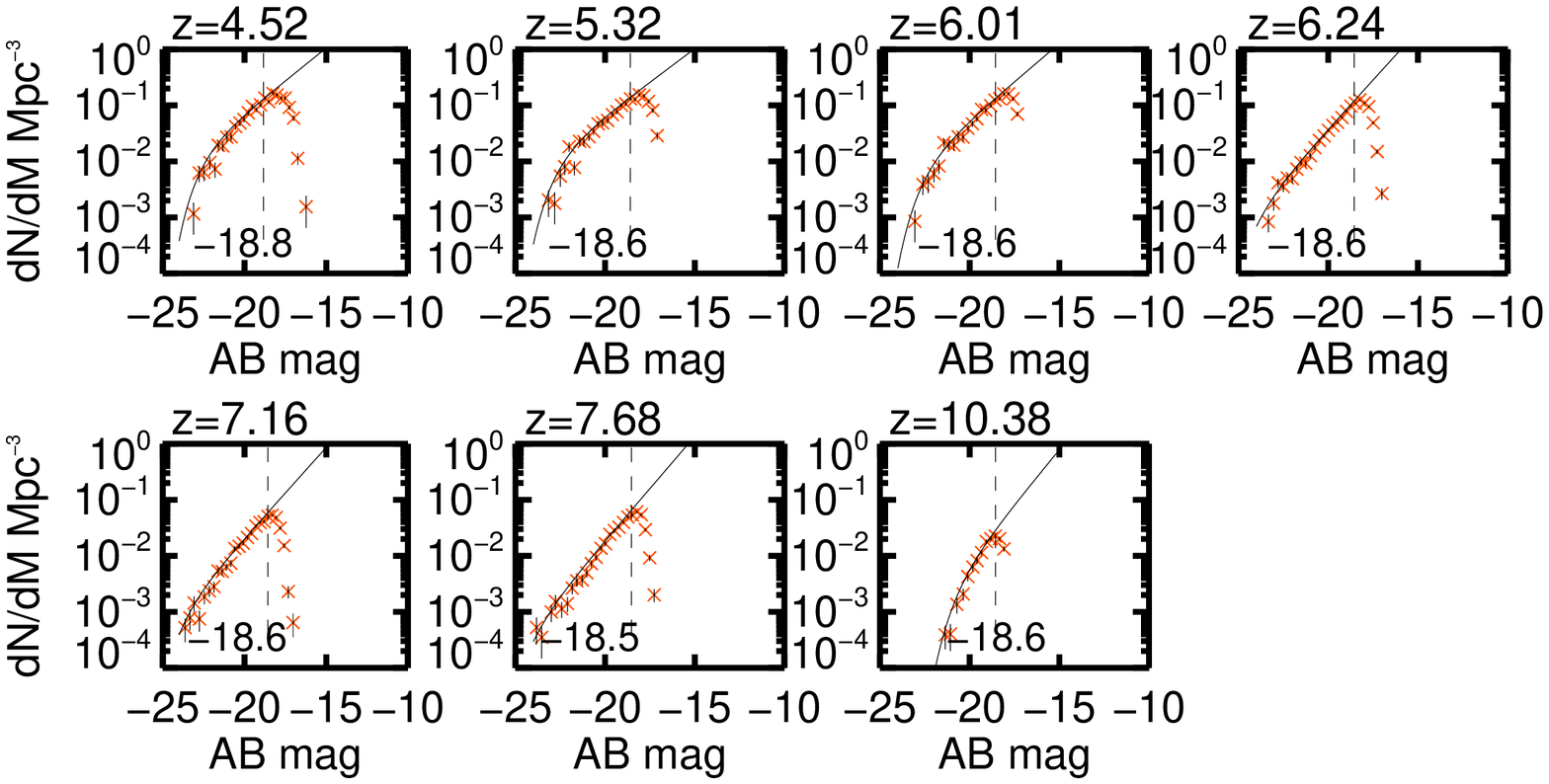}
\epsscale{.95}
\plotone {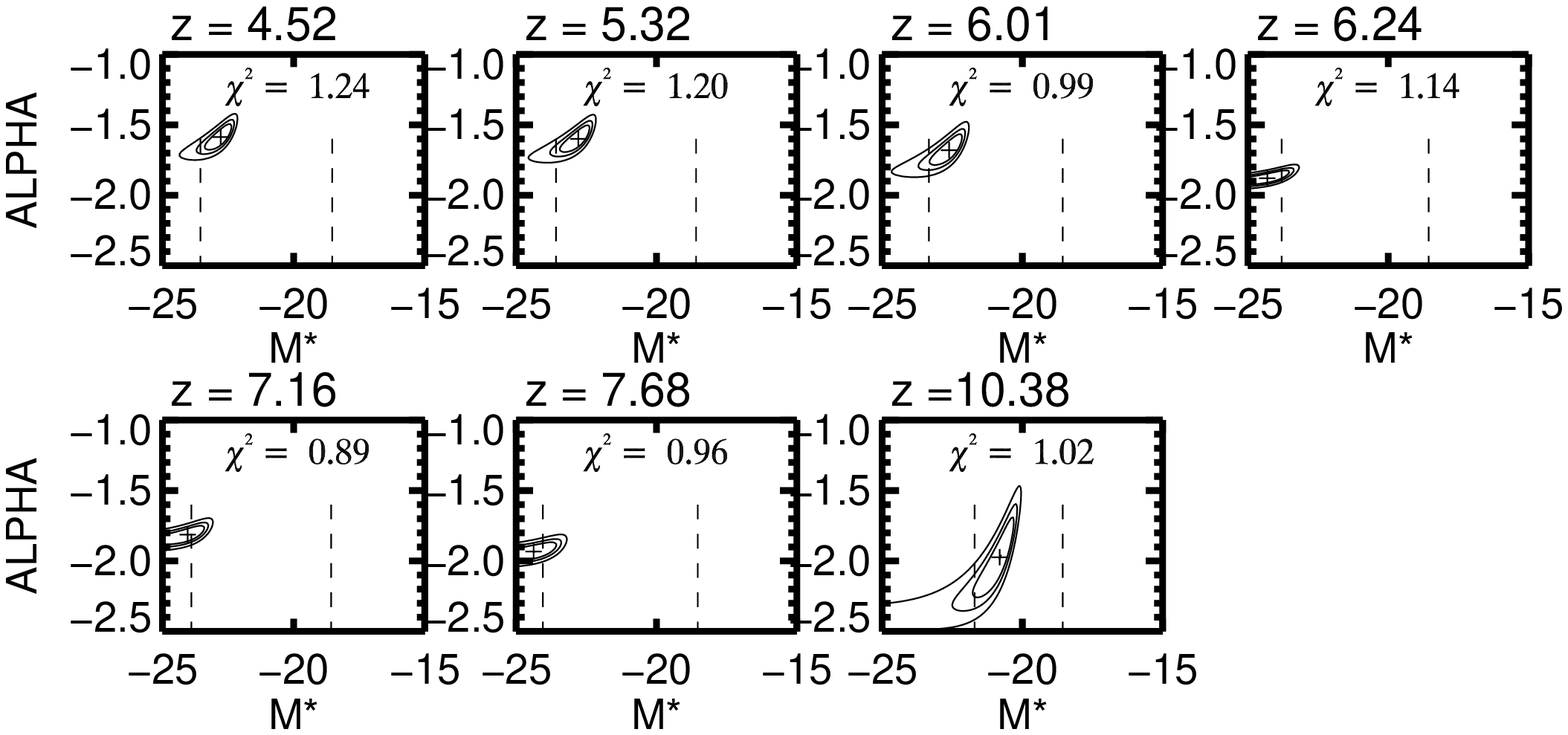}
\caption{
The Luminosity Function ${dN/dM/Mpc^3}$ vs. AB-mag of simulations 
\emph{without feedback} winds but \emph{with added sky-background noise}
 of $\mathrm{\approx22.6}$ AB-mag${/arcsec^2 }$. 
The lower panels are the confidence-region plots of best-fit Schechter 
functions in the $\alpha$--${M^*}$ parameter space. 
The contours indicate 68\%, 90\%, and 99\% confidence levels. 
They are at numerically simulated redshifts of 4.5, 5.3, 6.01, 6.24, 7.16, 
7.68, and 10.38. The vertical dotted lines in the upper LF plots indicate 
the faint-magnitude cutoffs used when obtaining a chi-square fit to  the
Schechter function with a reduced chi-square value of no more than 1.25, 
and $M_{AB}$ brighter than $ \simeq -18.5$ mag, to exclude the incompleteness 
region of the LF. Note the effect of the sky-background noise on the 
incompleteness of the LF, and how this outweighs the non-Schechter turnover in 
the LF of the simulated data without sky-BG at magnitudes fainter than 
$M_{AB} \simeq -17.5 $ mag (see Fig. 2).}
\end{center}
\end{figure}

\subsection{Effects on faint-end LF slope $\alpha$ from Feedback due to `Winds'}
In Figs. 6 \& 7, we examine the fits of the LFs to a Schechter function with
feedback `winds' enabled. A smaller dependence of $\alpha$ on redshift 
is now seen in the model results (Figs. 9 \& 10) when the `winds' feedback are 
enabled. We note that these simulations cover only a higher redshift 
range, due to the limited computing time available for running these more 
expensive simulations. 

We do not get a good fit to the observed data of Hathi et al. (2010)
over their observed redshift range (see Fig. 9). However, when  
comparisons are made over the redshift range $6.0 < z < 10.4$, the 
$\alpha$--redshift dependence slope of the `no-winds' and `winds' cases are 
very similar (Fig. 11). For the best fit Schechter function of the `winds' 
feedback case over the redshift range 6.0 $<$ z $<$ 10.4, we find:
\begin{eqnarray}
\mathrm{ \alpha = -1.54 \pm 0.25 - ( 0.06 \pm 0.04) \ z}  \ \ \ \ \\
\textrm{ [with Winds and No Sky-background] } \nonumber
\end{eqnarray}
We note that the slope of this result compares well with the `no-winds' case 
(Eq.(7)), i.e., the $\alpha$(z) curves are parallel, but somewhat displaced
in $\alpha$--space.

When we add the effects of dust extinction, PSF, and sky--background as
described in $\S{5.3}$ (Fig. 11), we obtain the best-fit of the LFs to
a Schechter function over the redshift range 6.0 $<$ z $<$ 10.4:
\begin{eqnarray}
\mathrm{ \alpha = -1.84 \pm 0.44 - ( 0.04 \pm 0.07) \ z}  \ \ \ \ \ \ \\
\textrm{ [with Winds, Sky-background, Dust, and PSF] } \nonumber
\end{eqnarray}

We note the apparent lessening of the $\alpha(z)$ dependence alluded to 
previously. This is not totally unexpected.  One would expect a power-law 
dependence of the faint--end slope ${\alpha \simeq -2}$ 
in the initial galaxy development, due to the primordial power spectrum
which predicts $\alpha \equiv -2$.  As galaxy assembly evolution proceeds, 
we see a reduction in the steepness of $\alpha$, perhaps due to mergers, as 
modeled by the merger-tree calculations of Khochfar et al. (2007).

We note that Eq.(15) is very close to the `no-winds' simulation of Eq.(11) 
and (13).  At the higher redshift of $z \approx$ 10.4, we see a slightly more 
negative value of $\alpha$, although these are still close to 
$\alpha \simeq -2$.  There is a suggestion of $\alpha$ leveling out to a range 
of ${\alpha \simeq -2 }$ to ${-2.1}$, inspecting the data at redshifts higher 
than ${z \simeq 7.7}$, although more simulation data points are needed to see 
a definite trend.  Again, this is not unexpected from analytical theory. 
Also, we note an apparent overall steepening in $\alpha$ at redshifts 
$z > 6$ when the feedback `winds' mechanism is enabled in the simulation. 
This may be due to the time delay in star-formation introduced by the 
winds-feedback in the simulation.

We do not see much of a change in $M^*$ as a function of $z$. Finkelstein
et al.(2014, 2015) also found similar results, which they attributed to a
lack of dust extinction in high redshift galaxies. 
Here, as previously described, the feedback from the combined energy of 
supernovae imparts a velocity to the gas in the model. If the gas resolution 
could be improved by several orders of magnitude, this may lead to more 
realistic gas outflows. 

To summarize, the winds mechanism --- as implemented here --- does not seem 
to affect the evolution of $M^*$ as much as it seems to affect the $\alpha$ 
evolution. Performing additional simulation runs with different `winds' 
parameter settings --- which was limited due to computing time constraints --- 
may help to resolve this issue. Also, a finer resolution, such as `zooming' 
in on a part of the simulation where star-formation is active, and restarting 
the simulation from just prior to this would aid in discerning the effect of 
mass resolution.

Jaacks et al. (2012) ran different simulations at different mass
resolution scales and simulation cube sizes to cover the bright, medium and 
faint-end of the LF scale.  Our simulation volume lies between their medium and 
faint-end simulations. We are close to their mass resolution at the faint-end, 
where they had gas particles of mass ${1.91 \times 10^{5}h^{-1}\ }$M$_{\odot}$ 
since they used only ${2 \times 400^3}$ particles. However, for their 
bright-end they used ${2 \times 600^3}$ particles of DM and gas in a simulation
box size of ${100  h^{-1}}\ $Mpc. This results in a volume nearly 200 times 
larger than ours, permitting them to detect objects two orders of magnitude 
rarer than in our simulation which enabled them to explore the bright-end 
better. Jaacks et al. (2012) also reported problems with the break 
at the bright-end which is discussed more in $\S{6}$.

We note that over the range of the model parameters and the physics modeled,
similar slopes are seen in the $\alpha-z$ parameter space. 
The faint-end LF slope $\alpha$ is seen to become steeper with increasing 
redshift, and in some cases, parallel tracks somewhat offset in $\alpha$ are 
seen. So, the effect of winds seems to be to overall steepen the $\alpha$-
values somewhat, but to not change the slope of the $\alpha$-z relation 
drastically.

We also note that, while improved somewhat, we still see an excess of
objects compared with the observations. This is discussed more in $\S{6}$.

\begin{figure}
\begin{center}
\epsscale{0.98}
\plotone {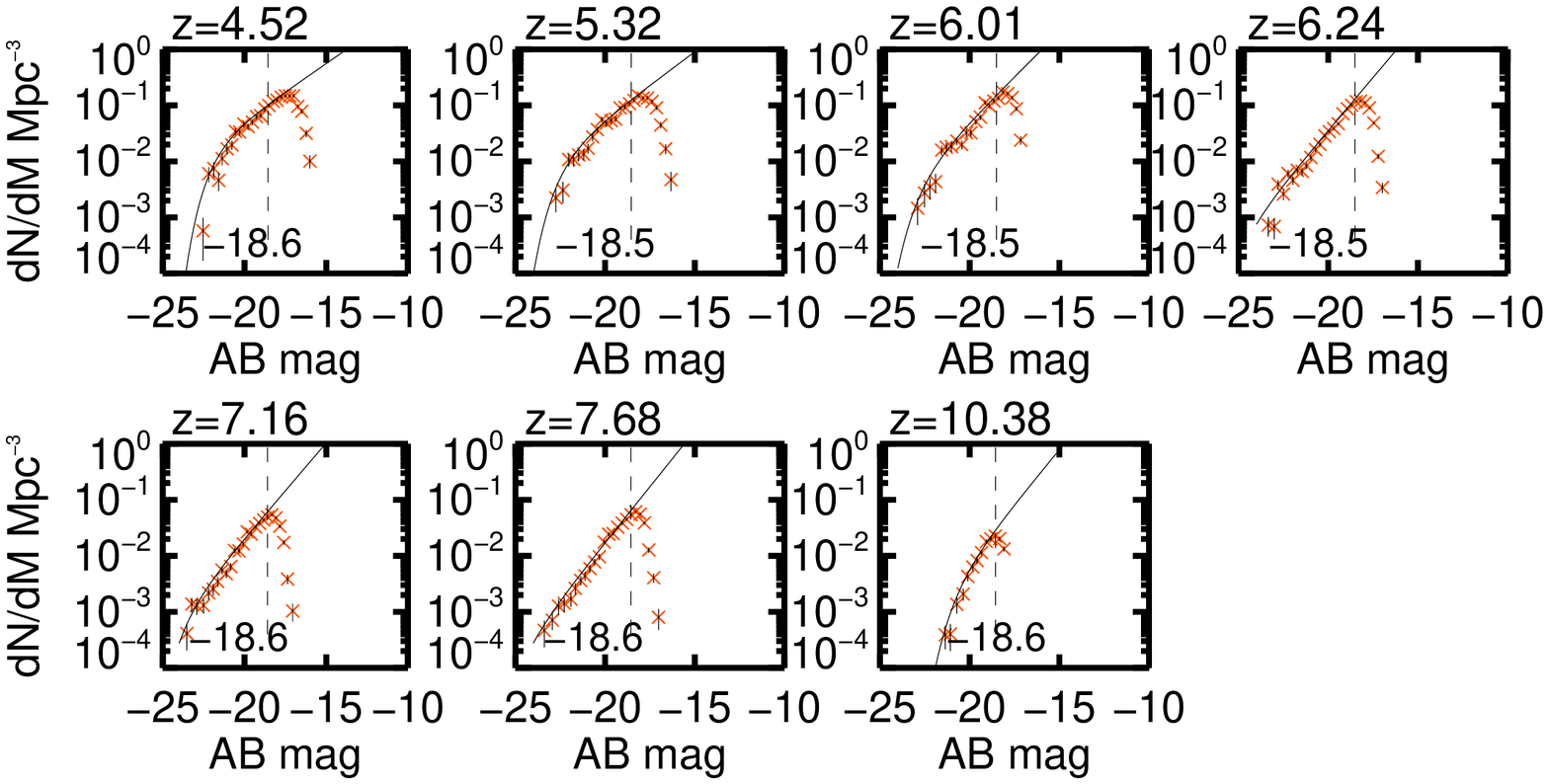}
\epsscale{0.95}
\plotone {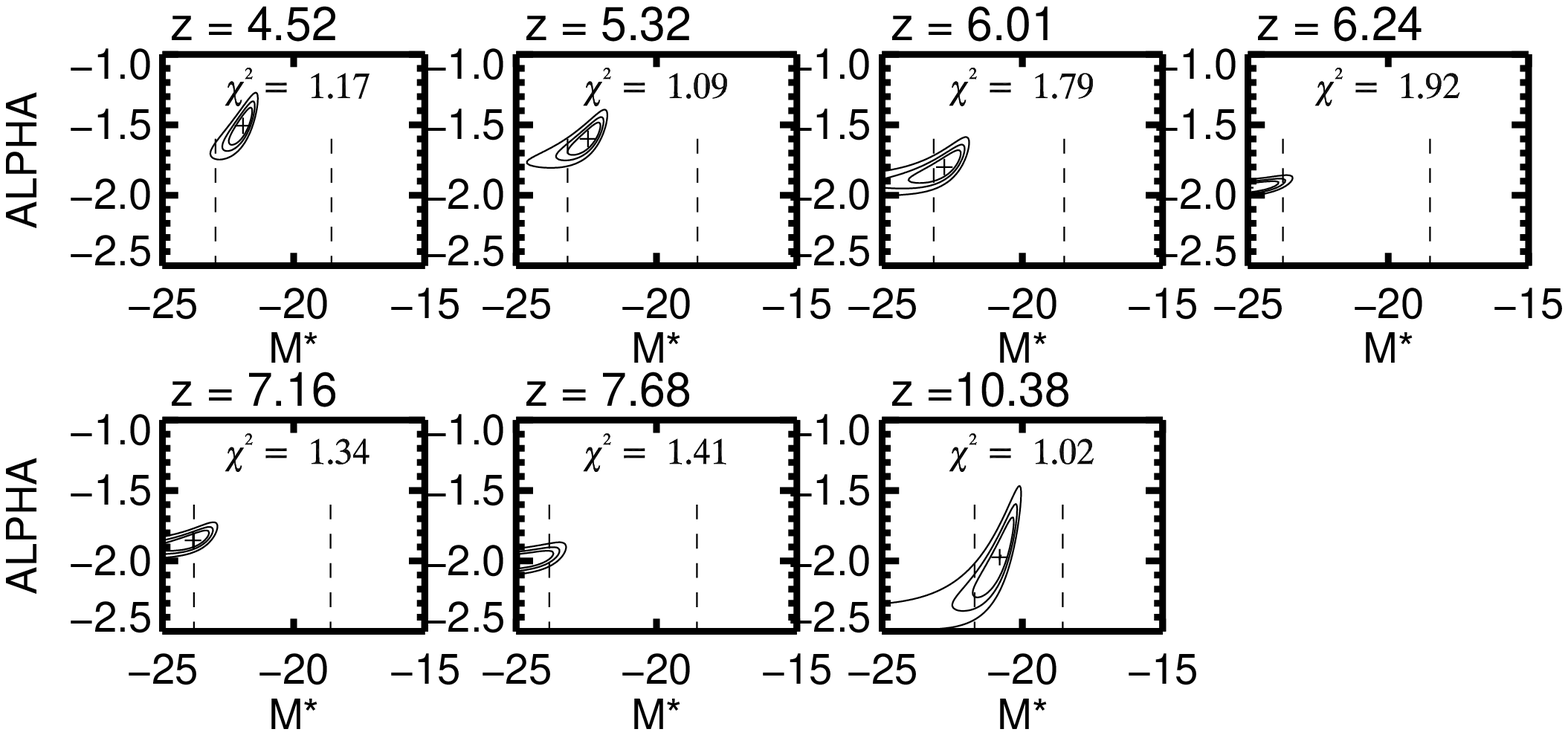}
\caption{The Luminosity Function ${dN/dM/Mpc^3}$ vs. AB-mag of simulations
\emph{without feedback} winds but \emph{with added sky-background noise and
with UV dust extinction}. The lower panels are the confidence-region 
plots of best-fit Schechter functions in the $\alpha$--${M^*}$ parameter space.
The contours indicate 68\%, 90\%, and 99\% confidence levels. 
They are at numerically simulated redshifts of 4.5, 5.3, 6.01, 6.24, 7.16, 
7.68, and 10.38. The vertical dotted lines in the upper LF plots indicate 
the faint-magnitude cutoffs used when obtaining a chi-square best-fit 
at luminosities brighter than $M_{AB}\simeq -18.5$ mag, as in the preceding 
plots. 
}
\end{center}
\end{figure}

\begin{figure}
\begin{center}
\epsscale{0.98}
\plotone {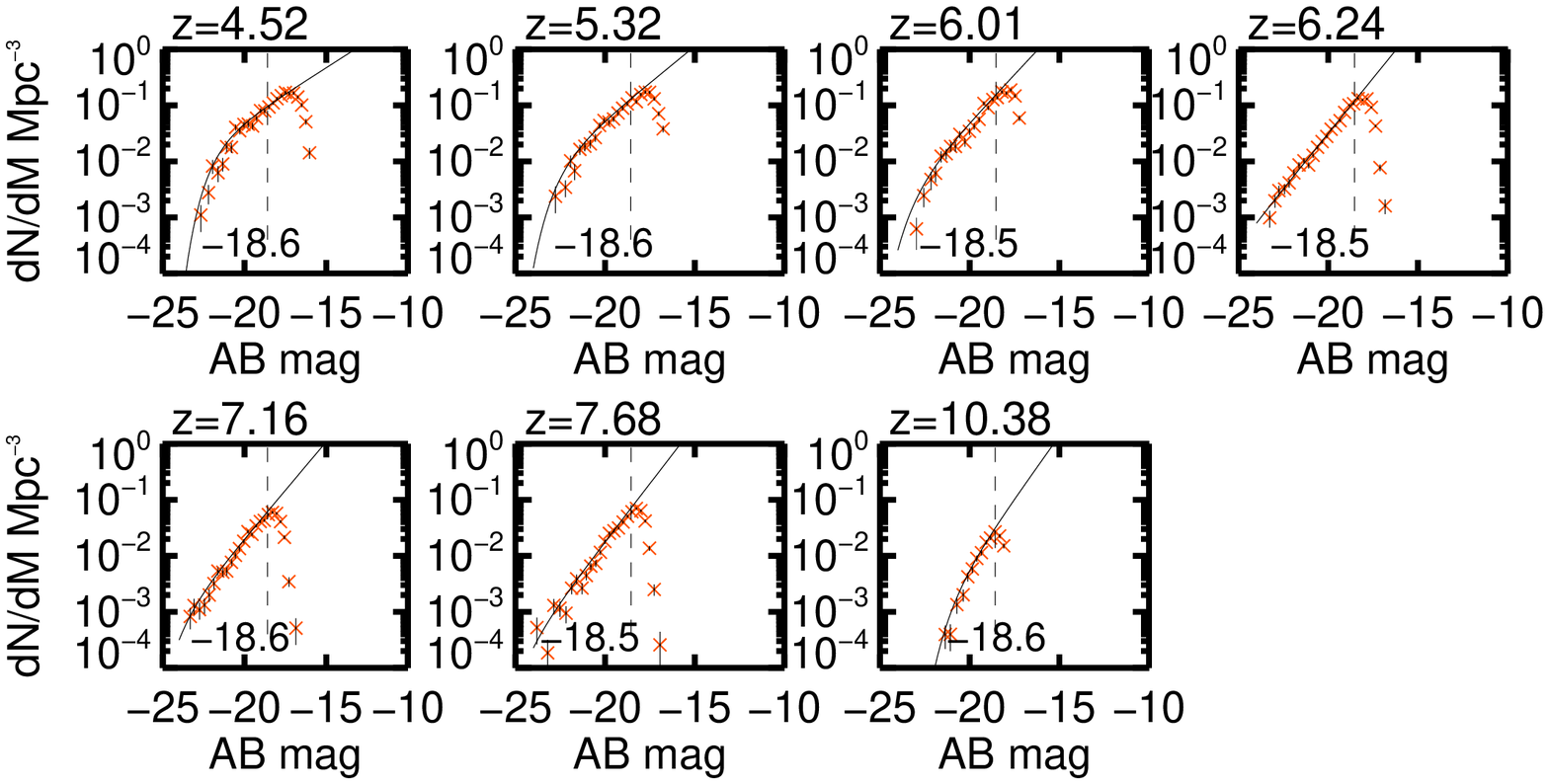}
\epsscale{0.95}
\plotone {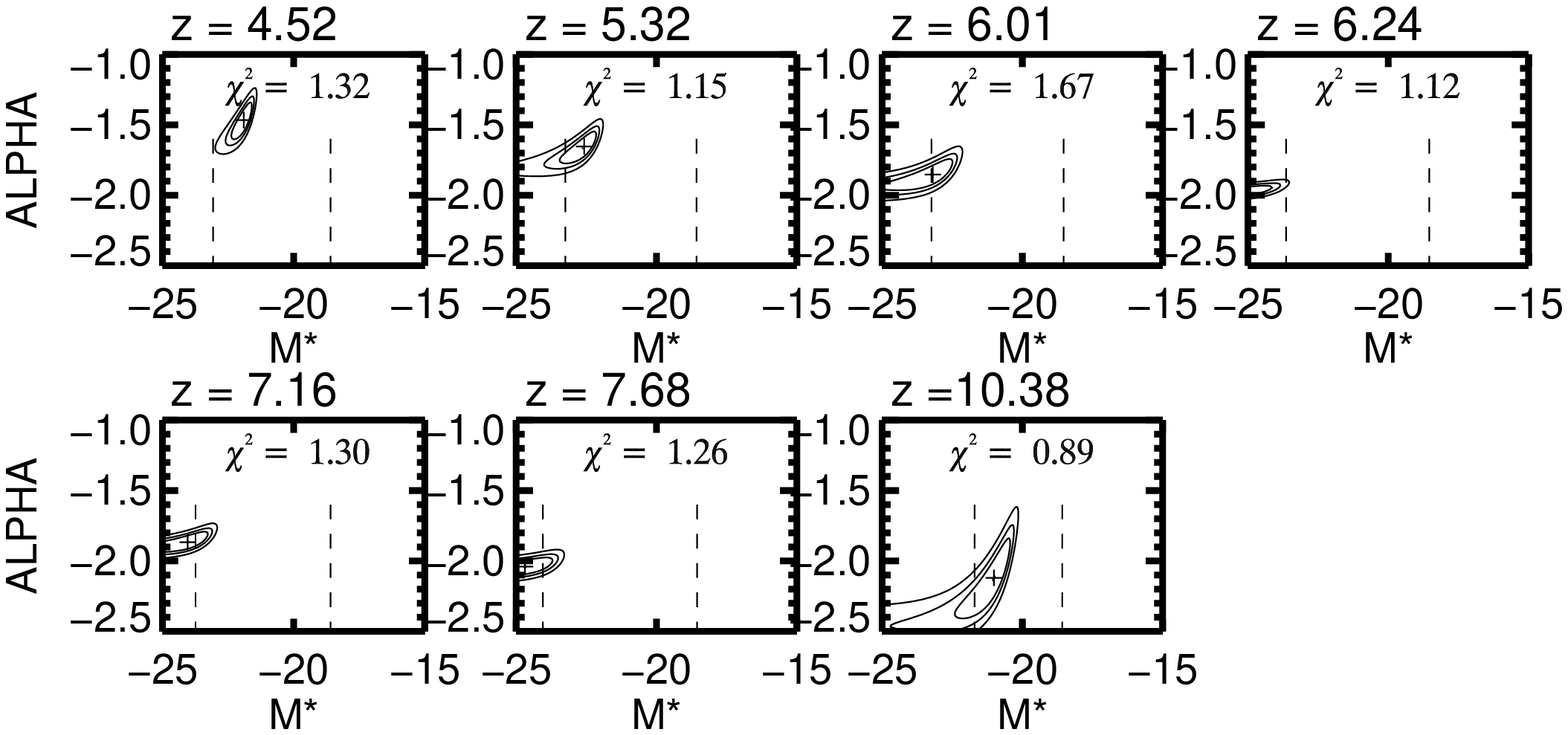}
\caption{The Luminosity Function ${dN/dM/Mpc^3}$ vs. AB-mag of simulations
\emph{without feedback} winds but \emph{with added sky-background noise and
with UV dust extinction} and convolved with a PSF of \emph{0.15" FWHM}. 
The lower panels are the confidence-region plots of the
best-fit Schechter functions in the $\alpha$--${M^*}$ parameter space.
The contours indicate 68\%, 90\%, and 99\% confidence levels. 
They are at numerically simulated redshifts of 4.5, 5.3, 6.01, 6.24, 7.16, 
7.68, and 10.38. The vertical dotted lines in the upper LF plots indicate 
the faint-magnitude cutoffs, used when obtaining a chi-square best-fit 
at luminosities brighter than $M_{AB}\simeq -18.5$ mag, as in the preceding 
plots. 
}
\end{center}
\end{figure}

\begin{figure}[t]
\begin{center}
\epsscale{0.98}
\plotone {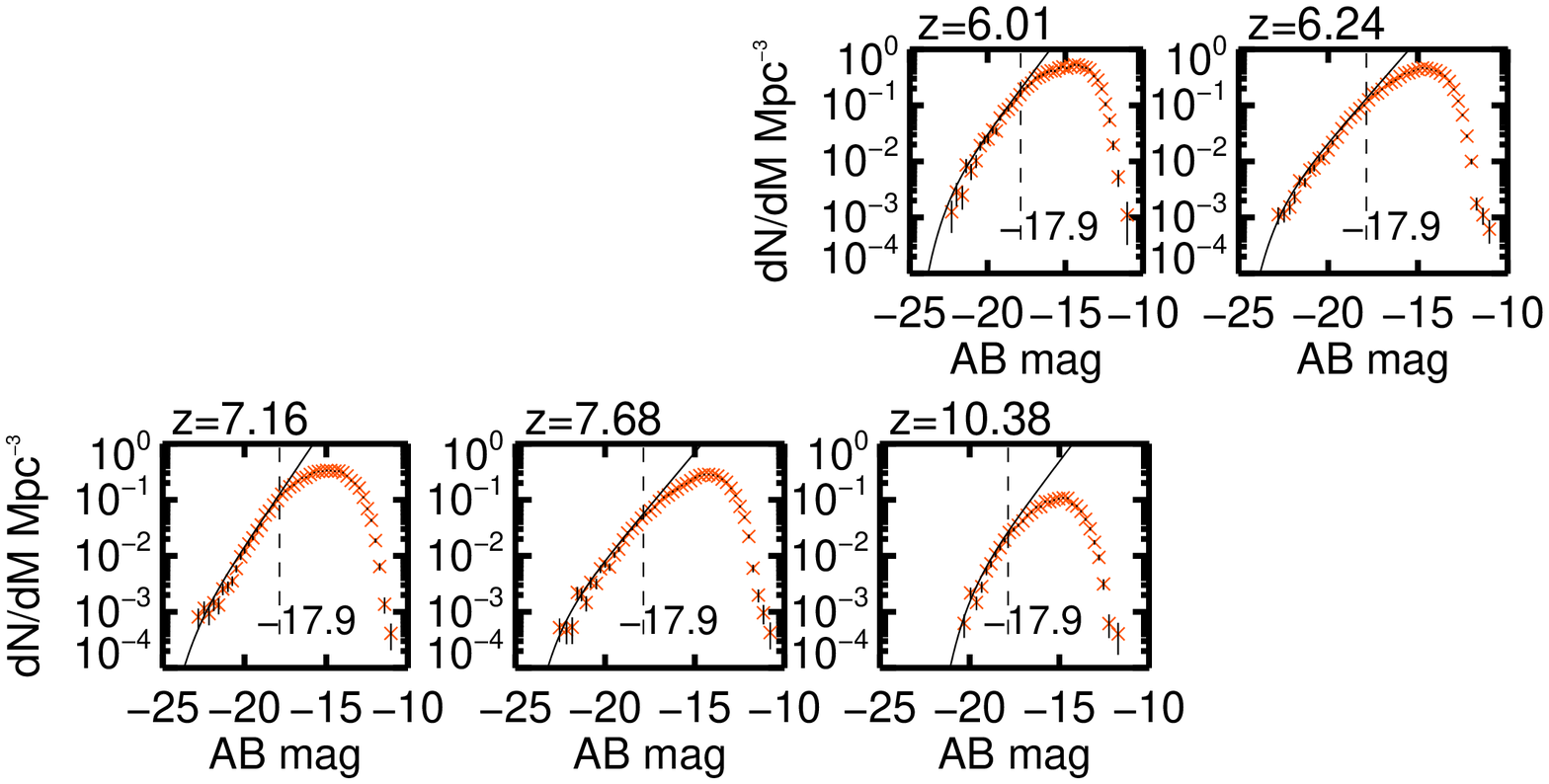}
\epsscale{0.95}
\plotone {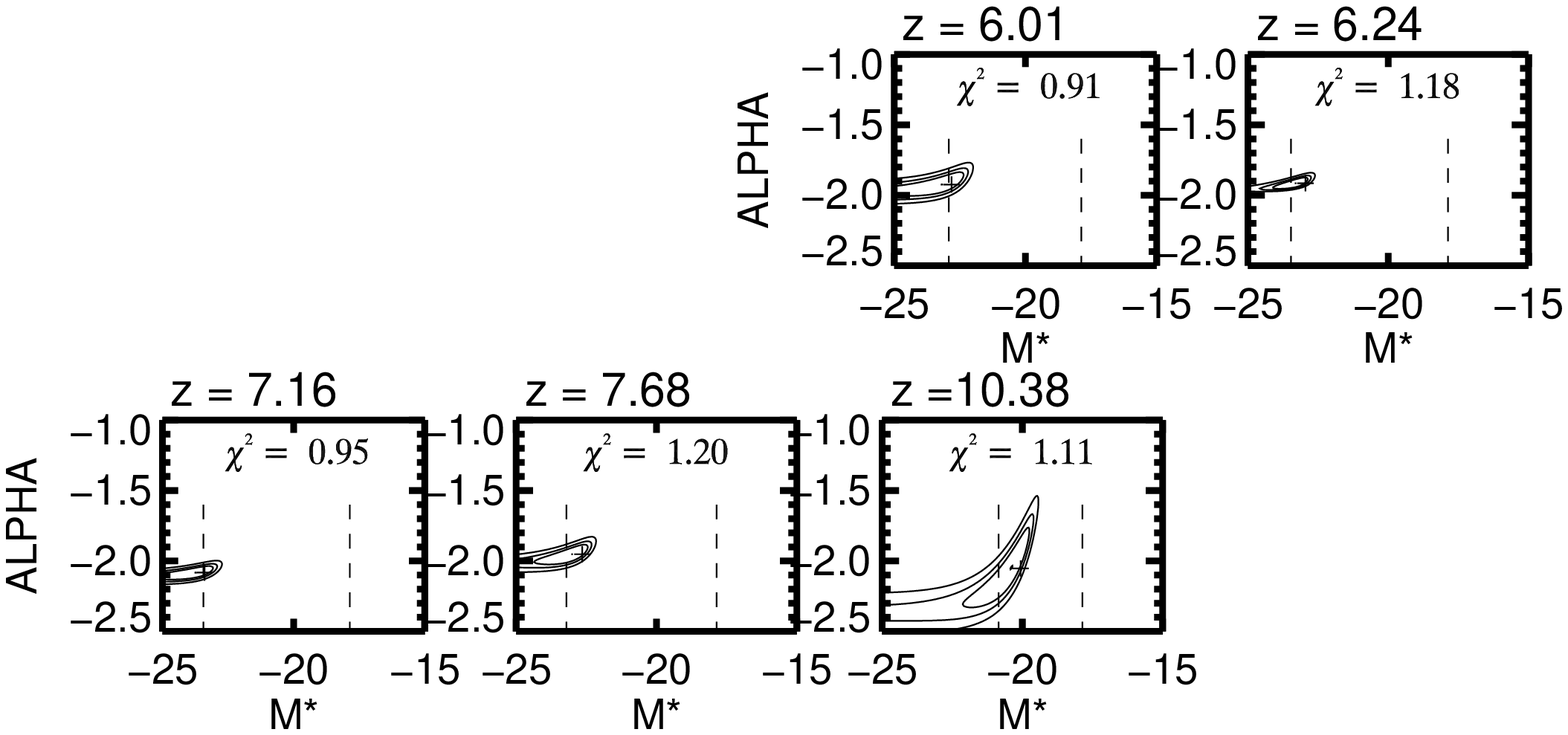}
\caption{The Luminosity Function ${dN/dM/Mpc^3}$ vs. AB-mag of simulations 
\emph{with feedback winds} and confidence-region panels of best-fit Schechter 
functions in the $\alpha$--${M^*}$ parameter space. The contours indicate 
68\%, 90\%, and 99\% confidence levels. These are from images created without 
adding sky-background noise. They are at numerically simulated redshifts of 
6.01, 6.24, 7.16, 7.68, and 10.38. The vertical dotted lines in the upper LF 
plots indicate the faint-magnitude cutoffs, used when obtaining chi-square fits
to the Schechter function with reduced chi-square value of no more than 1.2, 
and a luminosity brighter than ${M_{AB} \simeq -17.5 }$ mag.
The missing lower panels are due to the much more expensive simulation,
resulting in not being able to continue it to z ${\simeq4-5}$ within the
available computing time.}
\end{center}
\end{figure}

\begin{figure}[t]
\begin{center}
\epsscale{0.97}
\plotone {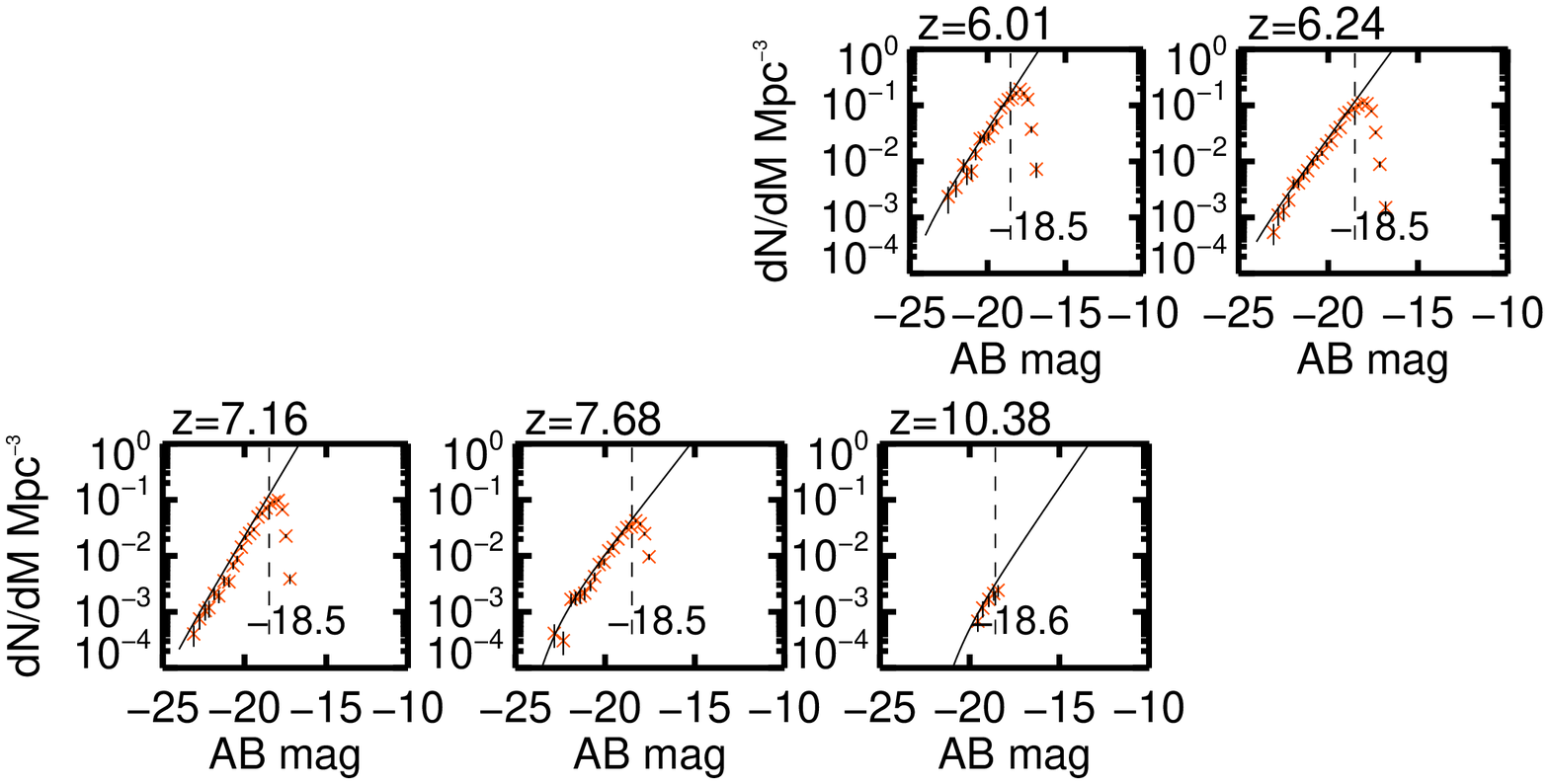}
\epsscale{0.95}
\plotone {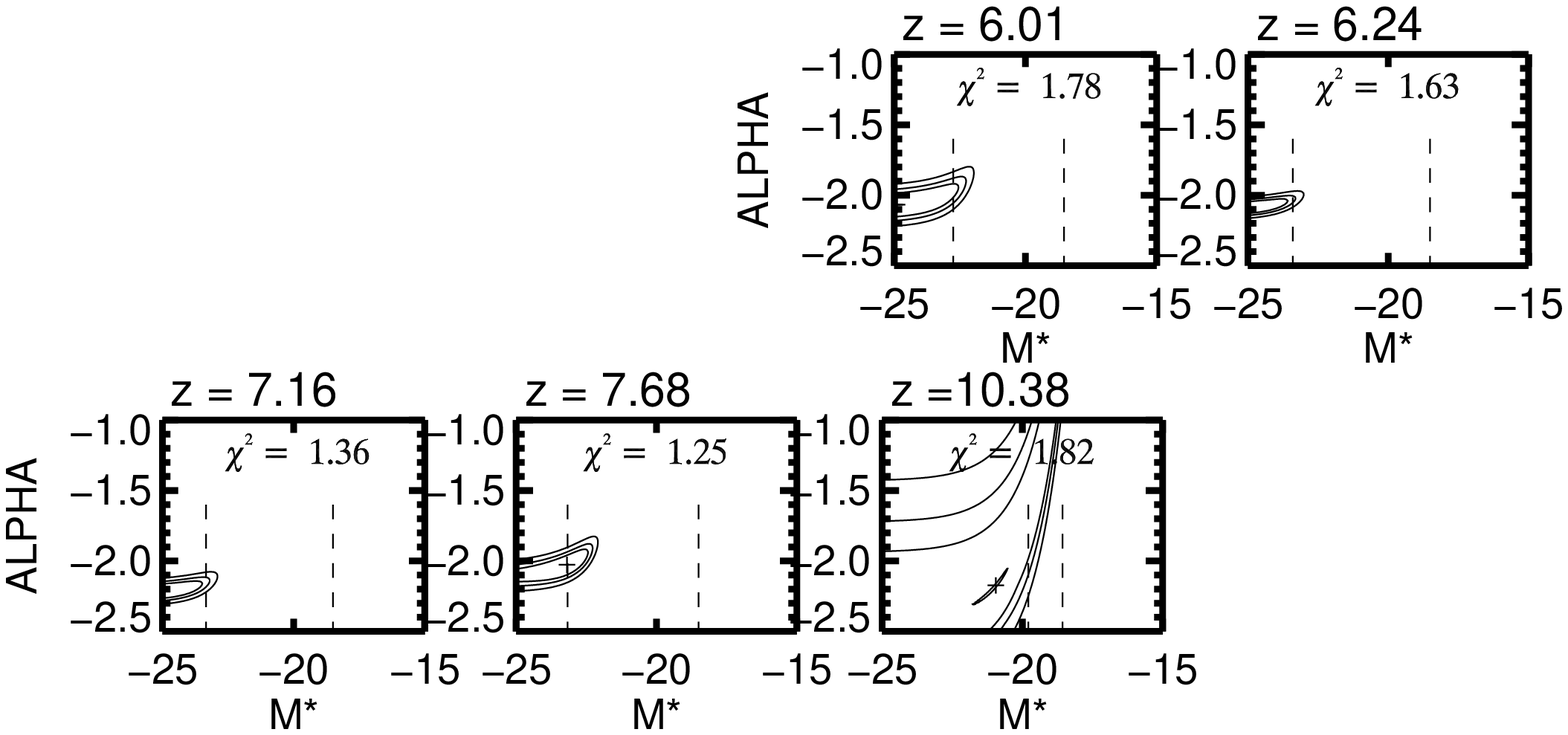}
\caption{The Luminosity Function ${dN/dM/Mpc^3}$ vs. AB-mag of simulations 
\emph{with feedback winds} and \emph{with sky-background, PSF, and dust}  
and confidence-region panels of the best-fit Schechter 
functions in the $\alpha$--${M^*}$ parameter space. The contours indicate 
68\%, 90\%, and 99\% confidence levels. These are from images created without 
adding sky-background noise. They are at numerically simulated redshifts of 
6.01, 6.24, 7.16 and 7.68.
The vertical dotted lines in the upper LF plots indicate the faint-magnitude 
cutoffs, used when obtaining a best-fit chi-square fit to the 
Schechter function with a luminosity brighter than ${M_{AB} \simeq -18.5 }$ 
mag.}
\end{center}
\end{figure}

\begin{figure}
\begin{center}
\epsscale{0.90}
\plotone {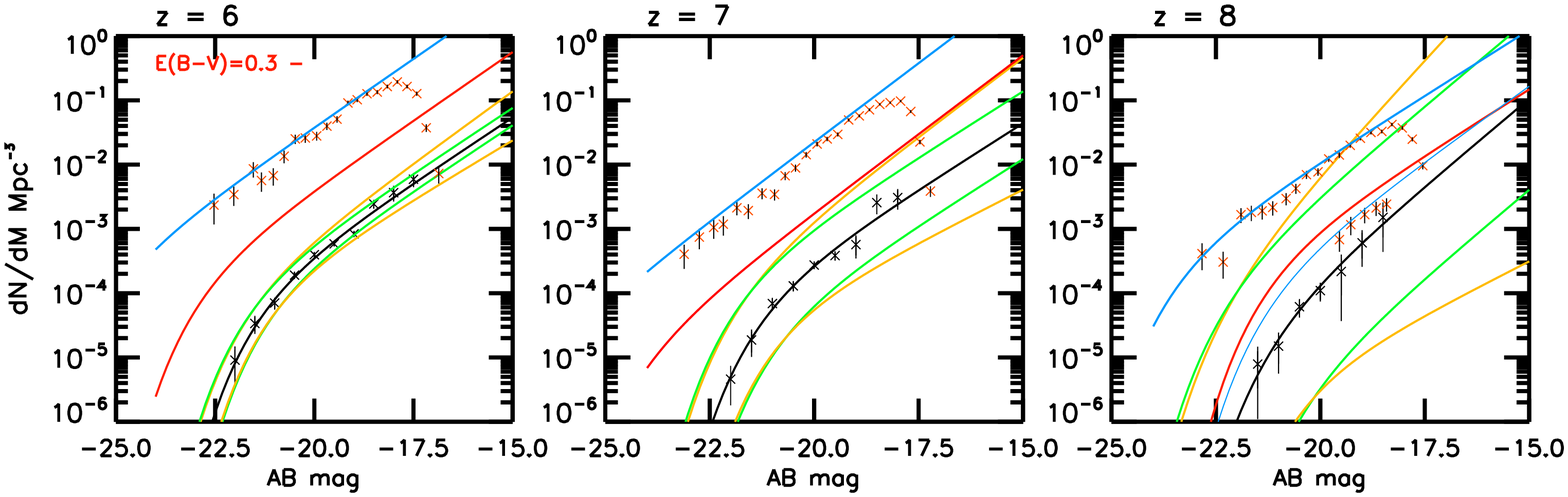}
\caption{The simulated LFs --- red x's --- vs. the observed LFs --- black 
	x's --- at redshifts 6, 7, and 8 from Finkelstein et al. (2014).
	The simulated LFs are \emph{with feedback winds} and 
	\emph{with sky-background, PSF, and dust}.
	The vertical black bars indicate the error estimates for each 
	magnitude bin.
	The blue curves indicate the best-fit Schechter functions for the 
	simulated LFs and the black curves indicate the best-fit Schechter 
	functions for the observed LFs. The green curves indicate 1--$\sigma$
	errors in M* and $\phi^{*}$ for the observed LFs and the orange curves
	including also the 1--$\sigma$ errors in the faint-end slope $\alpha$
	for the observed LFs.  The red curves indicate the effects of
	using a dust extinction on the simulated LF Schechter function
	best-fit based on an E(B-V)=0.3. The simulated redshifts are at 6.01,
	7.16, 7.68, and 10.38, respectively. Both of the simulated LFs at
	redshifts 7.68 and 10.38 are included on the rightmost plot with
	the observed LF at redshift 8. There is no dust extinction
	for the simulated redshift 10.38.}
\end{center}
\end{figure}

\subsection{Comparison of the Simulated LFs with Observation}

The simulated LFs at redshifts 6.01, 7.16, and 7.68 are compared to the
observed redshifts at 6, 7, and 8 LFs from Finkelstein et al. (2014) in
Fig. 8. Since Finkelstein et al. (2014) did not have an LF at redshift 10, we
include the simulated redshift 10.38 LF with the redshift 8 observed LF.
It will be noted that our volume densities are well in excess of the 
Finkelstein et al. (2014) densities, except for the simulated redshift 10.38. 
We have previously discussed that this is, in part due to our not selecting the
dust extinction which $\emph{creates a best fit}$ of the simulated LF to 
observed LFs. We include in the plots the effect of choosing an E(B-V)=0.30
on the Schechter fit, as given by the red curves in Fig. 8.
We also show the error curves for the Schechter function best-fits
to the observed LFs.  We see that increased extinction 
allows our Schechter fits to be very close to the Schechter fits of the 
observed LFs, when including the 1-$\sigma$ errors on $M^{*}$ and $\phi^{*}$ 
at redshift 7 and within those errors at redshift 8.

We also note that --- while at very different redshifts ---
that our simulated LF at redshift 10.38 lies very close to the observed LF
at redshift 8. 
A great increase is seen in the simulated LF density from redshift 10.38 to 
7.68, and a gradual increase at lower redshifts.
We note that the evolution in the faint-end slope $\alpha$ of the the LF
is similar in our simulated LFs and the observed LFs. 
Also, our smaller survey volume is seen to affect the ability of the simulation
to probe the bright-end of the LF, compared with the observed LFs.

\begin{figure} 
\epsscale{1.05}
\plotone {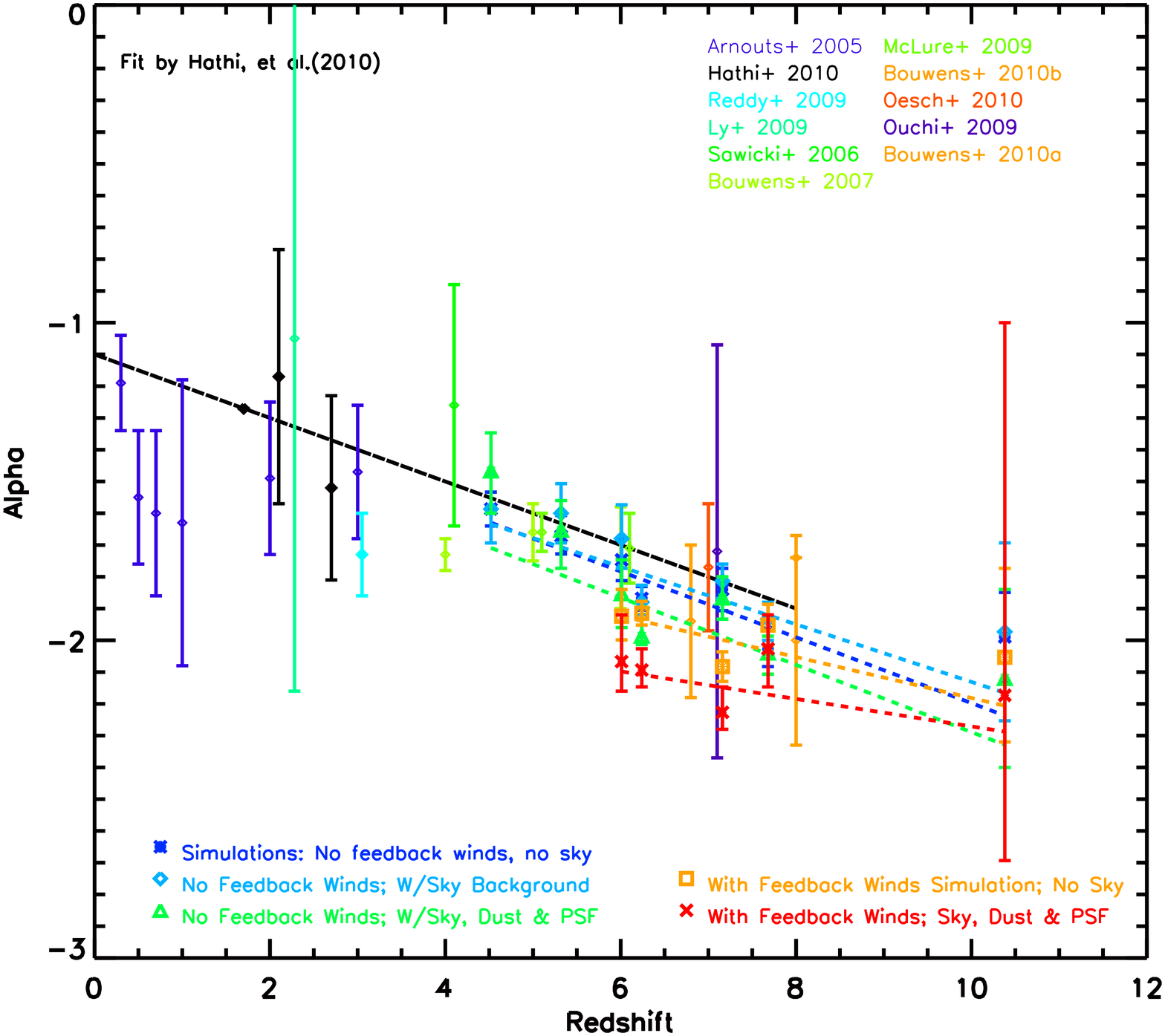}
\caption{The LF faint--end slope $\alpha$ vs. redshift from Hathi et al. (2010)
and references therein, with a best-fit of $\alpha (z) = -1.10 - 0.10 z $,
plus simulations \emph{without} feedback winds and  \emph{without} 
simulated sky-background [Eq.(6)] and \emph{with} simulated sky-background 
[Eq.(8)]. Also included are simulations \emph{with feedback winds} [Eq. (12)]
and with simulated dust absorption and instrumental PSF.} 

\end{figure}
%
\begin{figure} 
\epsscale{1.05}
\plotone {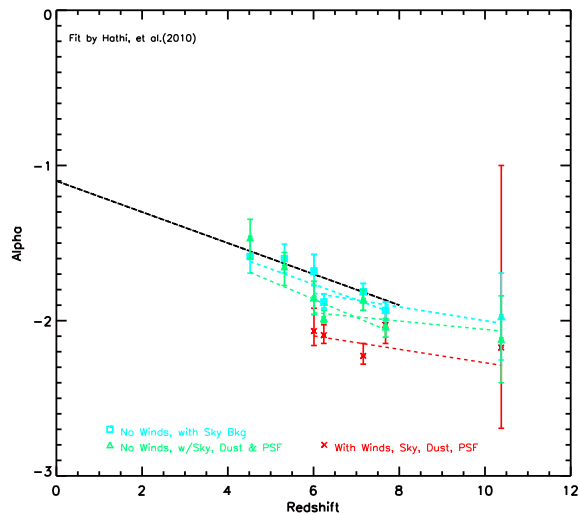}
\caption{The LF faint--end slope $\alpha$ vs. redshift from Hathi et al. (2010)
(Eq. (5)), compared with simulations \emph{without} feedback ``winds'' 
 and  \emph{with} sky-background (Eq. (10) and Eq. (8)), and \emph{with} PSF 
emulation and \emph{with} simulated dust-extinction, and \emph{with} feedback 
``winds'' over redshift ranges
$ 4.5 < z < 7.7 $ and $ 6.0 < z < 10.4$, using $\chi ^2 _3$ statistics (see $\S 5$).}
\end{figure}
\begin{figure} 
\epsscale{1.05}
\plotone {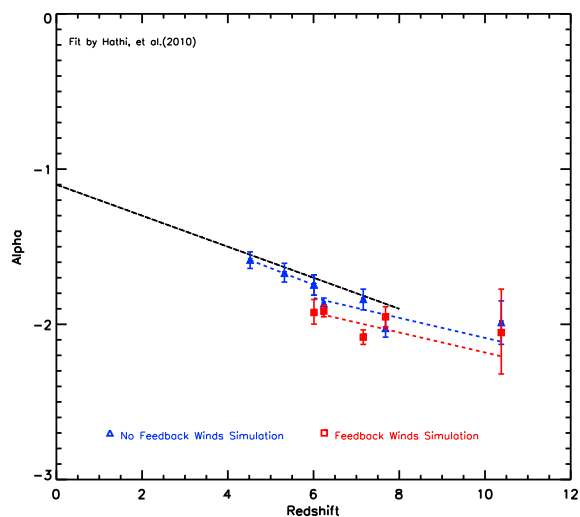}
\caption{The LF faint--end slope $\alpha$ vs. redshift from Hathi et al. (2010)
[Eq. (5)], plus simulations \emph{w/o} and \emph{with feedback} ``winds'', 
fitted over different redshift ranges, $6>z>4.5$ and $z>6$ (Eq. (7)) for basic 
non-feedback simulation, and $10.4>z>6$ for the simulation \emph{with feedback}
``winds'' case (Eq. 12), using $\chi ^2_3$ statistics (see $\S 5$).}
 
\end{figure}

\section{Discussion}  

%
Comparing our results with the observational fits of 
Hathi, et al. (2010) and references therein to the faint-end LF-slope 
$\alpha$ and the evolution of $\alpha$ with redshift, we find very similar 
results. However, there appears to be a change in the $\alpha (z)$ 
dependence at higher redshifts in our simulated data. 
Salvaterra et al. (2011) found similar results,
with their simulated LF faint-end slope having a nearly constant value of 
$\alpha = -2.0$ over redshifts 5 --- 10.

We note that the characteristic magnitude, $M^*$, vs. redshift does 
not appear to correspond well with the actual observations 
(e.g., Hathi et al. 2010, Bouwens, et al. 2011). 
The LF function may be better fit by a power--law than the
Schechter function at  $ M_{AB} \leq -16$ mag in our simulation results. 
This was also the conclusion of Bouwens, et al. (2007) and Bowler et al. 
(2014), regarding similar studies of galaxy formation. However, our results
may be due to limitations on our simulation, discussed below.

As we previously noted, the value of $M^*$ is generally outside --- or 
near the boundary  --- of the high-end of the magnitude range in 
the simulated data. Thus, $M^*$ may just be an indication of where the 
best--fit begins to depart from the Schechter function due to the limited 
simulation volume. Thus, it could be argued that $M^*$ is not reliably 
constrained by the currently simulated data, and that any apparent $M^*$ 
dependence on redshift is due mainly to the changes in the simulated data 
magnitude range with redshift. Jaacks et al. (2012) also noted this lack 
of a distinct break in the bright--end of the LF in their simulations. 
We note that they used a simulation at the bright--end with a much larger 
volume (see $\S{5.6}$), hence the survey volume limitations
may well not be the only cause. As noted in other work, simulations with 
AGN feedback may improve the break at the bright--end of the LF. Closer 
inspection of our LF curves shows some break at the very bright--end, 
indicating that even when including basic physics, there appears to be some 
mechanism for departure of the LF from power--law behavior.

Ultimately, the discrepancy with observed Schechter functions may be due
to several reasons. First, the $M^*$ break in the observations may be 
largely the result of feedback processes in the real universe, which are 
lacking in our initial model implementation. In real galaxies, as larger 
and hence more luminous amounts of gas and stars collect and form 
in the dark matter potential wells, more of the gas is ejected by feedback 
mechanisms, limiting the growth of these and in more luminous and massive 
galaxies also outflows from (weak) AGN. Without these mechanisms, one would 
just see a continuation of the power--law to brighter fluxes, as seen here. 
In order to test this, an attempt was made to include feedback in the form 
of `winds', which was not activated in the earlier model runs. The results 
of those models are seen in the Figs. 4 and 7, denoted as including `winds.'

One interesting feature in the model results is the break in the LF around
$M_{AB} \simeq -18$ mag and the rapid drop-off at fainter magnitudes (seen in
Fig. 2), that occurs even before including sky-background effects.  Note,  
due to sky-background induced incompleteness effects, that this feature is 
not seen --- or only hinted at --- in the sky-noise added simulated data in 
Fig. 3.  While this effect may be due to mass-resolution artifacts at the 
limiting mass of ${\approx 10^6 {M}_{\odot}}$, we note that this effect is 
seen at luminosities two orders of magnitude greater than the limiting 
magnitude of $M_{AB} = -12$ to $-10$ mag, assuming a constant M/L ratio.  
This same effect may also be seen in the simulations of Jaacks et al. (2012).  
This is of interest since, while hidden from observation due to the actual
sky-background, it could affect hierarchical formation by impacting the number 
of low-mass objects. This flux regime would be more accessible to future space 
missions, such as JWST.

We have previously noted that the number densities per unit volume are much
higher than observed, which may be due to the current treatment of 
feedback, which may not be treated with the appropriate resolution at this 
time. The difficulties of modeling feedback, and its importance to galaxy 
formation was also discussed by SH03 and others (e.g., Dayal et al. 2014).
Since the LF curve plots number density vs. luminosity, a slight decrease 
in the luminosity would reduce the discrepancy without altering the faint-end 
LF slope, which can be achieved by including the effects of dust extinction. 
In fact, as was described in $\S{3.6}$, some authors (e.g., Jaacks et al. 2012 
and Shimizu et al. 2014) selected their dust extinction values on the basis 
of achieving the best-fit to their LF to observed values.  We note that if we 
increase the dust extinction by several mags we can much improve the fit of 
our number densities to observed values.

While our initial objective was to study the LF faint-end, in order to 
explore the bright--end of the LF better --- and better simulate the 
characteristic magnitude $M^*$ --- we need to increase the simulation volume 
to include more luminous and therefore more  rare objects.
One way would be to run different volume and mass resolution simulations 
as in Jaacks et al. (2012) and combine them, as discussed in $\S{5.4}$. 
Another way would be to run an initial simulation at lower resolution, but 
a larger volume, and detect and model those objects of interest over this 
larger volume only, but at higher resolution.

\section{Summary}  
We have predicted the high-redshift galaxy UV LF using galaxy catalogs 
created by SExtractor from images derived from SSPs derived from a 
cosmological hydrodynamic simulation, with star-formation modeled via 
BC03 SSP SED models. We also added the Zodaical sky-background and additional 
feedback `winds' physics to investigate their effect on the UV LF.

We find close agreement with observed results (Hathi et al. 2010) for the 
faint-end slope $\alpha$ of the UV LF derived from our models.
We find little impact from considering winds and sky-background on these
observed results over a redshift range from ${z \simeq 4.5}$ to 
${z \simeq 10.4}$.  
That is, including the Zodi sky still predicts the same $\alpha$(z) relation
to within the errors, but to a brighter M$_{AB}$ limit.
We also see a similar evolution of $\alpha $ with 
redshift z, compared to the observed results (Hathi et al. 2010): 
${d \alpha (z) /d(z) \simeq -0.10 }$. Over this redshift range, we found 
$\alpha$ to increase from $\alpha$ $\simeq -1.5$ at $z \simeq 4.5$ to  
$\alpha \simeq -2.0$ at $z \simeq 10.4$. A slight flattening of the slope 
$\alpha (z)$ is seen at redshifts $z  > 6$ without the added sky-background: 
${d \alpha (z) /d(z) \simeq -0.06 \pm 0.03}$. With the sky-background 
included, we find a slightly less steep evolution, but still within the 
1--$\sigma$ errors.

The bright end of the luminosity function does not show the characteristic
magnitude $M^*$ reported in observations, though there is a hint of a drop off
in the LF density at the bright end, which may be due to the unavoidable volume
limitations of the simulation. As discussed in $\S{6}$, this effect has been 
reported by others (e.g., Jaacks et al. 2012), who had included large survey
volumes in their simulations, hence this volume limitation may not be the 
only factor. Feedback winds included in the simulation had surprisingly
little impact on the LF. This could be a resolution issue.  It is worth
noting that the winds include a delay in star-formation, which may impact
the LF density $\Phi(M)$.

We also show the possible effects that observational constraints, such as the 
Zodiacal background, may have on the ability to observe actual properties of 
galaxy formation. This is especially seen at the faint-end of the LF, due to
incompleteness effects from the added sky-background noise. 
We find a non--Schechter fit to the luminosity function in the regime 
fainter than M$_{AB} \approx -17$ mag in the model without sky-background.
This is the regime that is very difficult to access by direct observations. 
It needs to be investigated further if this is due to some artifact, 
or lack of certain physics in the model. If this result were confirmed to be 
real in future simulations, it might have implications for galaxy evolution 
through mergers, since there would be fewer building blocks at fainter 
magnitudes than a simple extrapolation of the Schechter law would predict.

\acknowledgments
We thank James Rhoads, Hwihyun Kim,Seth Cohen, and Rolf Jansen for helpful 
discussions and comments. We are grateful to the ASU Advanced Computing 
Center (A2C2) for making available computing time for the simulations.
RAW acknowledges support from NASA/JWST Grants NAG5-12460 and NNX14AN10G.

\end{document}